\documentclass[journal, twoside]{IEEEtran}
\usepackage{graphicx,amsmath,amssymb,color}
\usepackage{verbatim}
\usepackage[noadjust]{cite}
\usepackage{array}
\usepackage{cases}
\usepackage{url}
\usepackage{multirow}
\newtheorem{theorem}{Theorem}[section]
\newtheorem{proposition}[theorem]{Proposition}
\newtheorem{corollary}[theorem]{Corollary}
\newtheorem{lemma}[theorem]{Lemma}

\begin{document}
\title{Bounds on Separating Redundancy of Linear Codes and Rates of X-Codes}
\author{Yu Tsunoda,~\IEEEmembership{Student Member,~IEEE}, Yuichiro Fujiwara,~\IEEEmembership{Member,~IEEE}, Hana Ando, and Peter Vandendriessche%
\thanks{This paper was presented in part at the 2017 IEEE
International Symposium on Information Theory, Aachen, Germany.}%
\thanks{Y. Tsunoda is with the Graduate School of Science and Engineering, Chiba University, 1-33 Yayoi-Cho Inage-Ku, Chiba 263-8522, Japan
{(email: yu.tsunoda@chiba-u.jp)}.}%
\thanks{Y. Fujiwara is with the Division of Mathematics and Informatics, Chiba University, 1-33 Yayoi-Cho Inage-Ku, Chiba 263-8522, Japan
{(email: yuichiro.fujiwara@chiba-u.jp)}.}%
\thanks{H. Ando is with the Digital Platform Division, Nomura Research Institute, Tower N, 1-5-15 Kiba, Koto-ku, Tokyo 135-0042, Japan
{(email: h-ando@nri.co.jp)}.}%
\thanks{P. Vandendriessche is with the Department of Mathematics, Ghent University, Krijgslaan 281 - S22, 9000 Ghent, Belgium
{(email: peter.vandendriessche@ugent.be)}.}%
\thanks{This work was supported by JSPS KAKENHI Grant Number JP18J20466 (Y.T.), JSPS KAKENHI Grant Numbers JP15H06086 and JP17K12638 (Y.F.), and FWO (P.V.).
The fourth author is support by a postdoctoral fellowship of the Research Foundation - Flanders (FWO).}%
}

\markboth{Preprint typeset in IEEE Transactions style}%
{Tsunoda \MakeLowercase{\textit{et al.}}: Bounds on separating redundancy of linear codes and rates of X-codes}


\maketitle

\begin{abstract}
An error-erasure channel is a simple noise model that introduces both errors and erasures.
While the two types of errors can be corrected simultaneously with error-correcting codes,
it is also known that any linear code allows for first correcting errors and then erasures in two-step decoding.
In particular, a carefully designed parity-check matrix not only allows for separating erasures from errors but also makes it possible to efficiently correct erasures.
The separating redundancy of a linear code is the number of parity-check equations in a smallest parity-check matrix that has the required property for this error-erasure separation.
In a sense, it is a parameter of a linear code that represents the minimum overhead for efficiently separating erasures from errors.
While several bounds on separating redundancy are known, there still remains a wide gap between upper and lower bounds except for a few limited cases.
In this paper, using probabilistic combinatorics and design theory, we improve both upper and lower bounds on separating redundancy.
We also show a relation between parity-check matrices for error-erasure separation and special matrices, called X-codes, for data compaction circuits in VLSI testing.
This leads to an exponentially improved bound on the size of an optimal X-code.
\end{abstract}


\IEEEpeerreviewmaketitle

\section{Introduction}
\IEEEPARstart{D}{iscrete} error-erasure channels with input alphabet $\Gamma$ and output alphabet $\Gamma\cup\{e\}$, where $e \not\in \Gamma,$ are the simplest abstract models that combine two different types of fundamental channels.
For instance, the most elementary error-erasure channel is the natural combination of a binary symmetric channel and a binary erasure channel
in which each bit is independently either flipped to the other symbol with probability $p_{\text{error}}$, altered to $e$ with probability $p_{\text{erasure}}$,
or kept intact with probability $1-p_{\text{error}}-p_{\text{erasure}}$.
Such combined channels have been studied not only because they are natural from the purely theoretical viewpoint,
but also because they are reasonable models of noise in various scenarios.
Situations in which a combination of additive noise and loss of data occurs include wireless communications with cross-layer protocols \cite{Cheun:1991,Park:1998,Larzon:1999,Rath:2004,Karande:2005},
delay-sensitive optical communications \cite{McEliece:1981,Liva:2010} including some proposed deep-space communications systems \cite{Hemmati:2011,Cola:2011}, and magnetic and optical recoding \cite{Ha:2003,Yang:2004} among others.

There are various possible decoding algorithms that can handle simultaneous occurrences of errors and erasures.
Since the location of each erasure is known to the decoder, one straightforward approach is to assign a random symbol to each erasure and perform standard error correction.
With this method, when exactly $l$ symbols are erased, a naive decoder for a $\vert \Gamma\vert$-ary error-correcting code would try ${\vert \Gamma\vert}^l$ possible patterns to infer the original codeword.
Message-passing decoding methods such as belief propagation may also be used by assigning an appropriate likelihood to each erased symbol \cite{Richardson:2008}.

Recently, an interesting alternative approach was proposed in \cite{Abdel-Ghaffar:2013}, where errors and erasures are separately corrected by a linear code.
They revisited a classical strategy in which the decoder first corrects errors using the punctured code obtained by discarding the erased symbols
and then handles the erasures with the original linear code.
The innovative part of their idea is use of a carefully chosen parity-check matrix that allows for instantly providing parity-check matrices for appropriate punctured codes on demand while making efficient erasure correction possible by iterative decoding.

As we will formally define later, a parity-check matrix is called $l$-\textit{separating} if it admits this on-demand separation for all patterns of $l$ or fewer erasures.
It is not difficult to prove that any linear code of minimum distance $d$ has an $l$-separating parity-check matrix for any $l \leq d-1$ if there is no restriction on the number of redundant parity-check equations.
However, to minimize overhead for implementation and reduce decoding complexity, we would like our parity-check matrix to be as small as possible.

The $l$-\textit{separating redundancy} of a linear code $\mathcal{C}$ is the number of parity-check equations in a smallest $l$-separating parity-check matrix for $\mathcal{C}$.
While the $l$-separating redundancy of a linear code is important in the study of error-erasure separation,
it appears to be quite difficult to give a precise estimate, let alone determine the exact value.
Indeed, the precise values are known only for a few limited cases such as for the binary extended Hamming codes with $l=1$ \cite{Abdel-Ghaffar:2008}
and for maximum distance separable (MDS) codes with a few specific $l$ \cite{Abdel-Ghaffar:2013a}.
For linear codes in general, the current best upper and lower bounds are still far apart \cite{Liu:2018,Lie:2015,Abdel-Ghaffar:2013}.

The purpose of this paper is twofold.
Our first objective is to improve the estimate of the separating redundancy of a linear code through probabilistic combinatorics and design theory.
We refine both upper and lower bounds on $l$-separating redundancy that work for any linear code.
To more sharply bound the parameter from below, we improve the simple volume bound given in \cite{Abdel-Ghaffar:2013} to one of Sch\"{o}nheim type \cite{Schonheim:1964}.
We also give new strong upper bounds using probabilistic combinatorics.
In addition to these, a known upper bound based on design theory is refined.
While the design-theoretic approach is not as universally strong as the probabilistic one, it is shown to give a sharper estimate than any other known bound in some cases.
As will be illustrated by numerical computations, these results collectively make a meaningful improvement to our knowledge on separating redundancy.
As far as the authors are aware, this is the first substantial general progress towards closing the gap between upper and lower bounds since the introduction of separating redundancy.

The second but equally important objective of this paper is to relate the study of separating redundancy to that of compaction circuits for efficient testing of Very-large-scale integration (VLSI) chips.
Due to the extreme complexity of modern computer circuits, it is very expensive and certainly not a trivial task to check whether a given VLSI chip was flawlessly manufactured and is working correctly \cite{McCluskey:2003}.
\textit{X-codes} are the mathematical abstraction of a type of data compaction circuit that was invented to reduce the overall cost of this task \cite{Lumetta:2003,Mitra:2004,Mitra:2005,Fujiwara:2010}.
As we will briefly explain later, optimal X-codes are those that compress a certain kind of data for circuit testing as much as possible while retaining sufficient test quality.
Note that these codes are not related to the special MDS array codes which were introduced in \cite{Xu:1999} and also happen to be called X-codes.
We show a close relation between $l$-separating parity-check matrices and X-codes.
This connection is then exploited to exponentially improve the best known general bound on the size of optimal X-codes.

In the next section, we briefly review the concept of separating redundancy and known bounds.
Section \ref{sec:lower} gives our improved lower and upper bounds and explains how our approaches mathematically refine the previously known techniques from a general viewpoint.
Numerical examples in Section \ref{sec:tables} illustrate how our bounds compare against the known general bounds in specific cases.
The relationship between error-erasure separation and X-codes is studied in Section \ref{sec:Xcodes},
where a new bound for X-codes is also provided.
Section \ref{sec:conclusion} concludes this paper with some remarks.

\section{Preliminaries}\label{sec:pre}
In this section, we mathematically define separating redundancy and review known results.
While we also define most of the basic notions in coding theory we use, for more comprehensive treatments, we refer the reader to standard textbooks such as \cite{MacWilliams:1977,Roth:2006}.

As usual, an $[n,k,d]_q$ \textit{linear code} $\mathcal{C}$ of \textit{length} $n$, \textit{dimension} $k$, and \textit{minimum distance} $d$ over the finite field $\mathbb{F}_q$ of order $q$ is a $k$-dimensional subspace of the $n$-dimensional vector space $\mathbb{F}_q^n$ over $\mathbb{F}_q$ such that
$\min\{\operatorname{wt}(\boldsymbol{c}) \mid \boldsymbol{c} \in \mathcal{C}\setminus\{\boldsymbol{0}\}\} = d$,
where $\operatorname{wt}(\boldsymbol{a})$ for $\boldsymbol{a} \in \mathbb{F}_q^n$ is the Hamming weight of $\boldsymbol{a}$.
Each vector in $\mathcal{C}$ is a \textit{codeword} of the linear code.

The \textit{dual code} $\mathcal{C}^{\perp}$ of the linear code $\mathcal{C}$ is the Euclidean dual space of $\mathcal{C}$, that is,
$\mathcal{C}^{\perp} = \{\boldsymbol{c} \in \mathbb{F}_q^n \mid \boldsymbol{c}\cdot\boldsymbol{c}' = \boldsymbol{0} \text{\ for any\ } \boldsymbol{c}' \in \mathcal{C}\}$.
The \textit{dual distance} $d^{\perp}$ of $\mathcal{C}$ is the minimum distance of its dual code $\mathcal{C}^{\perp}$.
Regarding each element of $\mathbb{F}_q^n$ as an $n$-dimensional row vector over $\mathbb{F}_q$,
a \textit{parity-check matrix} $H$ for $\mathcal{C}$ is an $m \times n$ matrix over $\mathbb{F}_q$ whose rows span $\mathcal{C}^{\perp}$.
A \textit{supercode} $\mathcal{D}$ of $\mathcal{C}$ is a set $\mathcal{D} \subseteq \mathbb{F}_q^n$ such that $\mathcal{C} \subseteq \mathcal{D}$.
If $\mathcal{D}$ is also a linear code, its parity-check matrix consists of some, but not necessarily all, codewords of $\mathcal{C}^{\perp}$ and is of rank at most $n-k$ over $\mathbb{F}_q$.
If $\mathcal{D} \not= \mathcal{C}$, it is a \textit{proper} supercode.

We use the nonnegative integers less than $n$ to specify the coordinates of the vector space $\mathbb{F}_q^n$.
For a set $S \subset \{0,1,\dots,n-1\}$ of coordinates,
the \textit{punctured code} $\mathcal{C}_{\overline{S}}$ of the $[n,k,d]_q$ linear code $\mathcal{C}$ by $S$ is the $[n-\vert S\vert,k',d']_q$ linear code for some $k' \leq k$ and some $d' \leq d$ obtained by deleting the coordinates in $S$ from the codewords of $\mathcal{C}$.
In other words, $\mathcal{C}_{\overline{S}}$ is the linear code obtained by \textit{puncturing} $\mathcal{C}$ on $S$.

Assume that the sender transmitted a codeword of an $[n,k,d]_q$ linear code $\mathcal{C}$
and that the channel introduced to the codeword exactly $\vert S\vert$ erasures on the coordinates $S \subset \{0,1,\dots,n-1\}$ along with some errors elsewhere in the same $n$-symbol block.
While there can be multiple ways to overcome a combination of errors and erasures,
one of the simplest methods is to correct the errors by the punctured code $\mathcal{C}_{\overline{S}}$ and then the erasures by $\mathcal{C}$.
It is well known that for any fixed integer $z \geq 0$, this error-erasure separation strategy provides a decoding algorithm that is guaranteed to correct up to $x$ errors and up to $y$ erasures if $2x+y \leq d-1-z$
and declares detection of an anomaly if $z \geq 1$ and $d-z \leq 2x+y \leq d-1$ (see, for example, \cite{Roth:2006}).
It should be noted that this strategy may be able to correct more severe combinations of errors and erasures.
For instance, if we employ a reasonable but aggressive decoding method such as minimum distance decoding at the error correction step,
the decoder does not necessarily fail or give up decoding
even if the number of errors substantially exceeds half of the minimum distance of $\mathcal{C}_{\overline{S}}$.

For error-erasure separation to be effective, however, we need to be able to provide parity-check matrices for appropriate punctured codes without requiring unnecessarily long time or large memory space.
It is shown in \cite{Abdel-Ghaffar:2013} that this can be done with a special matrix that makes it possible to easily find a parity-check matrix for the relevant punctured code and doubles as one for the linear code for efficient erasure correction.

Let $H$ be a parity-check matrix for a given $[n,k,d]_q$ linear code $\mathcal{C}$.
Define $H(S)$ to be the submatrix of $H$ obtained by discarding all rows that contain a nonzero element in at least one coordinate in $S$ and deleting all columns corresponding to the coordinates in $S$.
The submatrix $H(S)$ is a parity-check matrix for a supercode of $\mathcal{C}_{\overline{S}}$.
The parity-check matrix $H$ is $S$-\textit{separating} if $H(S)$ is a parity-check matrix for $\mathcal{C}_{\overline{S}}$.
We call $H$ $l$-\textit{separating} if it is $S$-separating for any subset $S \subset \{0,1,\dots,n-1\}$ of cardinality less than or equal to $l$.

To give a concrete example, consider parity-check matrix
\[H =
\begin{pmatrix}
1 & 1 & 0 & 0 & 0 & 0 & 1 & 1\\
1 & 1 & 0 & 0 & 1 & 1 & 0 & 0\\
0 & 1 & 0 & 1 & 0 & 1 & 0 & 1\\
0 & 0 & 1 & 1 & 1 & 1 & 0 & 0\\
0 & 0 & 1 & 1 & 0 & 0 & 1 & 1\\
1 & 0 & 1 & 0 & 1 & 0 & 1 & 0
\end{pmatrix}
\]
for the $[8,4,4]_2$ extended Hamming code.
For $S = \{0,1\}$, we have
\[H(S) = 
\begin{pmatrix}
1 & 1 & 1 & 1 & 0 & 0\\
1 & 1 & 0 & 0 & 1 & 1
\end{pmatrix},
\]
which is a correct parity-check matrix for the $[6,4,2]_2$ linear code obtained by puncturing the extended Hamming code on the first two bits.
Hence, this particular $H$ is $\{0,1\}$-separating.
However, it is not $2$-separating because, for example, if $S = \{0,7\}$,
then we have
\[H(S) =
\begin{pmatrix}
0 & 1 & 1 & 1 & 1 & 0
\end{pmatrix},
\]
which does not have enough linearly independent rows to be a valid parity-check matrix for the punctured code obtained by deleting the first and last coordinates.

Roughly speaking, an $l$-separating parity-check matrix has valid parity-check matrices for all required punctured codes ready for use as conveniently stored submatrices under the assumption that at most $l$ erasures can happen.
Indeed, for any pattern of $l$ or fewer erasures, a parity-check matrix for the corresponding punctured code can be obtained by taking the rows that do not check any of the erased symbols and then throwing away the zeros at the erased positions.

Interestingly, if $l$ is smaller than the minimum distance $d$ of a given linear code $\mathcal{C}$,
it can be shown that for any set $S$ of coordinates of cardinality $\vert S \vert \leq l$, an $l$-separating parity-check matrix for $\mathcal{C}$ contains a row that has exactly one nonzero element in the coordinates in $S$ \cite{Abdel-Ghaffar:2013}.
In other words, for any pattern of $l$ or fewer erasures, it always contains a row that checks exactly one erased coordinate,
so that erasures can be corrected quickly one by one without solving a system of linear equations.

Note that the above property relates $l$-separating parity-check matrices to a well-known concept for efficient erasure correction.
A \textit{stopping set} $E$ in an $m \times n$ parity-check matrix $H$ for a linear code of length $n$ is a set of columns in $H$
such that the $m \times \vert E\vert$ submatrix of $H$ formed by the columns in $E$ does not contain a row of weight one \cite{Di:2002}.
This concept is important in modern coding theory because parity-check matrices that contain no small stopping sets admit very efficient iterative decoding over an erasure channel.
Indeed, stopping sets and related concepts have been studied extensively in coding theory (see, for example, \cite{Schwartz:2006,Hollmann:2007,Han:2007,Han:2008,Han:2008a,Weber:2008,Colbourn:2009}).
In the case of $l$-separating parity-check matrices, however, we assume that the received vector may contain not only erasures but also errors.

Trivially, for $l \geq 1$, an $l$-separating parity-check matrix is also $(l-1)$-separating.
It is also straightforward to see that if $\vert S\vert \leq d-1$, $H(S)$ is a parity-check matrix for $\mathcal{C}_{\overline{S}}$ if and only if $\operatorname{rank}(H(S)) = n-k-\vert S\vert$ over $\mathbb{F}_q$.
Note that if $\vert S\vert \geq d$, the dimension of $\mathcal{C}_{\overline{S}}$ can be less than $k$, in which case the erasure pattern that corresponds to $S$ cannot be corrected by $\mathcal{C}$.
Note also that if $S$ is the empty set, any parity-check matrix is $S$-separating.
Hence, in what follows, we follow the spirit of the original study of $l$-separating redundancy in \cite{Abdel-Ghaffar:2013} and exclude the trivial cases that $l = 0$ and that $l \geq d$.

In general, all else being equal, it is more desirable for an $l$-separating parity-check matrix to have fewer rows.
However, even for a small parity-check matrix $H$, it can be a daunting task to check for all possible $S$ with $\vert S\vert \leq l$ whether $H(S)$ is a valid parity-check matrix for $\mathcal{C}_{\overline{S}}$.
The following proposition makes it easier to verify that a given parity-check matrix is $l$-separating.
\begin{proposition}[\cite{Abdel-Ghaffar:2013}]\label{prop1}
Let $\mathcal{C}$ be an $[n,k,d]_q$ linear code.
Then, for any $l \leq \min\{d,n-k\}-1$, a parity-check matrix for $\mathcal{C}$ is $l$-separating if and only if it is $S$-separating for any $S \subset \{0,1,\dots,n-1\}$ of cardinality $l$.
\end{proposition}

The above proposition says that for virtually all cases, we only need to check the $\binom{n}{l}$ patterns of exactly $l$ erasures rather than all possible patterns of up to $l$ erasures.
The only exceptional case is $(d-1)$-separation of an $[n,k,d]_q$ linear code with $d = n-k+1$,
that is, $(d-1)$-separation of a \textit{maximum distance separable} (\textit{MDS}) code.
In this unique exceptional case, the situation is even simpler.
\begin{proposition}[\cite{Abdel-Ghaffar:2013}]\label{prop:MDS}
Any parity-check matrix for any $[n,k,n-k+1]_q$ linear code is $S$-separating for any $S \subset \{0, 1, \dots, n-1\}$ with $\vert S\vert = n-k$.
\end{proposition}

The above proposition implies that a $(d-2)$-separating parity-check matrix for an MDS code of minimum distance $d$ is automatically $(d-1)$-separating,
which is generally not the case with other linear codes.
Thus, for MDS codes of minimum distance $d$, we only need to consider the case $l \leq d-2$, where Proposition \ref{prop1} is applicable the same way as in any other linear code.

A crucial problem regarding the error-erasure separation method is how small an $l$-separating parity-check matrix can be.
The $l$-\textit{separating redundancy} $s_l(\mathcal{C})$ of an $[n,k,d]_q$ linear code $\mathcal{C}$ is the number of rows in a smallest $l$-separating parity-check matrix for $\mathcal{C}$.
Note that $s_{d-1}(\mathcal{C}) = s_{d-2}(\mathcal{C})$ for an MDS code $\mathcal{C}$ of minimum distance $d$ because
in this case a $(d-2)$-separating parity-check matrix is also $(d-1)$-separating by Proposition \ref{prop:MDS}.
For this reason, we focus on the case $l \leq \min\{d,n-k\}-1$,
that is, $l \leq d-2$ for an MDS code and $l \leq d-1$ for any other linear code.

As far as the authors are aware, the following is the only known nontrivial lower bound on separating redundancy for a general linear code.
\begin{theorem}[{\cite[Theorem 9]{Abdel-Ghaffar:2013}}]\label{lwb:volume}
Let $\mathcal{C}$ be an $[n,k,d]_q$ linear code with dual distance $d^{\perp}$.
For any $l \leq \min\{d,n-k\}-1$, the $l$-separating redundancy $s_l(\mathcal{C})$ satisfies
\[s_l(\mathcal{C}) \geq \frac{\binom{n}{l}(n-k-l)}{\binom{n-d^{\perp}}{l}}.\]
\end{theorem}

While the above bound can be proved by a simple volume argument, it is nonetheless achieved by some linear codes for some $l$ \cite{Abdel-Ghaffar:2008,Abdel-Ghaffar:2013a}.

There are several known general upper bounds on $s_l(\mathcal{C})$ that are of comparable usefulness.
Among them, the one based on the pigeonhole principle is often the sharpest, especially when the parameter $l$ is not too large and no structural information about $\mathcal{C}$ except its basic code parameters are available.

To describe the upper bound in a concise manner, we use the number $f_q(a,b)$ of $a \times b$ matrices over $\mathbb{F}_q$ of rank $b$ with $b \leq a$ that do not contain all-zero rows,
which was proved in \cite{Abdel-Ghaffar:2013} to be
\[f_q(a,b) = \sum_{i=0}^a(-1)^i\binom{a}{i}\prod_{j=0}^{b-1}(q^{a-i}-q^j).\]
\begin{theorem}[{\cite[Theorem 10]{Abdel-Ghaffar:2013}}]\label{upb:probabilistic}
Let $\mathcal{C}$ be an $[n,k,d]_q$ linear code.
For any $l \leq \min\{d,n-k\}-1$, the $l$-separating redundancy $s_l(\mathcal{C})$ is less than or equal to the minimum integer $t$ that satisfies
\[\sum_{i=0}^t\binom{t}{i}f_q(i,l)\frac{\prod_{j=0}^{n-k-l-1}(q^t-q^{i+j})}{\prod_{h=0}^{n-k-1}(q^t-q^h)} >1-\frac{1}{\binom{n}{l}}.\]
\end{theorem}

A weaker version of the above bound in closed form is also available in \cite{Abdel-Ghaffar:2013}.

The following is another general bound proved by extending the idea of \textit{generic erasure-correcting sets} \cite{Hollmann:2006,Hollmann:2007,Ahlswede:2012}.
\begin{theorem}[{\cite[Corollary 5]{Abdel-Ghaffar:2013}}]\label{upb:generic}
Let $\mathcal{C}$ be an $[n,k,d]_q$ linear code.
For any $l \leq \min\{d,n-k\}-1$, the $l$-separating redundancy $s_l(\mathcal{C})$ satisfies
\[s_l(\mathcal{C}) \leq \sum_{i=1}^{l+1}\binom{n-k}{i}(q-1)^{i-1}.\]
\end{theorem}

While the right-hand side of the inequality in the above theorem is quite large compared to the lower bound in Theorem \ref{lwb:volume}, it was proved by a constructive method and gives an upper bound in simple form.
In addition, it is relatively sharper when $l$ is large and can even beat the other known upper bounds.

The above general upper bounds depend on the alphabet size $q$ of a linear code, while the lower bound given in Theorem \ref{lwb:volume} does not.
Interestingly, there is also one known general upper bound that does not depend on $q$.

To describe the bound, we need a special combinatorial design.
In what follows, $\mathbb{N}$ represents the set of positive integers.
Let $n, \mu, l, \lambda \in \mathbb{N}$ be such that $n \geq \mu \geq l$.
An $l$-$(n,\mu,\lambda)$ \textit{covering} is a pair $(V, \mathcal{B})$ of a finite set $V$ of cardinality $n$ and a collection $\mathcal{B}$ of $\mu$-subsets of $V$
such that every $l$-subset of $V$ appears in at least $\lambda$ elements of $\mathcal{B}$.
The \textit{covering number} $C_{\lambda}(n,\mu,l)$ is the cardinality $\vert \mathcal{B}\vert$ of a smallest $\mathcal{B}$ such that there exists an $l$-$(n,\mu,\lambda)$ covering $(V, \mathcal{B})$.
For combinatorics on coverings and covering numbers, the interested reader is referred to \cite[Section VI-11]{:2007b} and references therein.
\begin{theorem}[{\cite[Corollary 3]{Abdel-Ghaffar:2013}}]\label{upb:covering}
Let $\mathcal{C}$ be an $[n,k,d]_q$ linear code.
For any $l \leq \min\{d,n-k\}-1$, define $L= \{i \in \mathbb{N} \mid l \leq i \leq \min\{d,n-k\}-1\}$.
The $l$-separating redundancy $s_l(\mathcal{C})$ satisfies
\[s_l(\mathcal{C}) \leq \min_{\mu \in L}\left\{(n-k-\mu)C_{1}(n,\mu,l)+\binom{n}{l}(\mu-l)\right\}.\]
\end{theorem}

The covering number $C_{1}(n,\mu,l)$ has been investigated extensively in combinatorics.
In particular, using a strong probabilistic method known as the \textit{R\"{o}dl nibble} \cite{Alon:2016}, it was shown that
$C_{1}(n,\mu,l)$ is asymptotically $\binom{n}{l}/\binom{\mu}{l}$.
\begin{theorem}[\cite{Rodl:1985}]
For fixed integers $2 \leq l \leq \mu$,
\[\frac{\binom{n}{l}}{\binom{\mu}{l}} \leq C_{1}(n,\mu,l) \leq (1+o(1))\frac{\binom{n}{l}}{\binom{\mu}{l}},\]
where the $o(1)$ term tends to zero as $n$ tends to infinity.
\end{theorem}

For a comprehensive list of results on bounds on and known exact values of $C_{1}(n,\mu,l)$, we refer the reader to \cite{:2007b} (see also \cite{Chee:2013,Barber:2016,Keevash:2018} for the more recent results not covered in the list).

There are several other bounds on separating redundancy that consider very special cases such as $1$-separation for cyclic codes as well as linear codes whose duals contain the vector of all ones.
One may also derive an upper bound for a linear code if its weight distribution is partially known.
For those specialized bounds, the interested reader is referred to \cite{Abdel-Ghaffar:2013,Lie:2015}.

\section{New bounds on separating redundancy}\label{sec:lower}
This section is divided into three subsections to present our bounds on separating redundancy in an organized manner.
Section \ref{subsec:lwb} refines the lower bound in Theorem \ref{lwb:volume}.
This new bound, Theorem \ref{thm:lwb}, is strictly stronger than the old bound in the sense that for any linear code, our version is at least as strong and quite often sharper.
Our probabilistic upper bounds are given in Section \ref{subsec:upb}.
Among these, our main bound is Theorem \ref{thm:probabilistic}, which is simpler and consistently gives a sharp upper bound.
Section \ref{subsec:coveringupb} proves a design-theoretic upper bound that improves Theorem \ref{upb:covering}.
The results on upper bounds from combinatorial designs are summarized in Corollary \ref{coro:upbcovering}.
While this is not universally strong, our numerical examples show that the design-theoretic bound occasionally gives the strongest known estimate for some parameters.

\subsection{Improved lower bound}\label{subsec:lwb}
While the lower bound in Theorem \ref{lwb:volume} is achievable in some cases, a simple observation shows that using the idea of a covering leads to a tighter bound.

We first generalize the concept of a covering $(V, \mathcal{B})$ by allowing the elements of $\mathcal{B}$ to have different sizes.
Let $n$, $\mu$, $l$, and $\lambda$ be positive integers and $K_\mu$ a finite set of positive integers such that $n \geq \mu \geq l$ and such that $\mu$ is the largest element in $K_\mu$.
An $l$-$(n,K_\mu,\lambda)$ \textit{generalized covering} $(V, \mathcal{B})$ is a pair $(V, \mathcal{B})$ of a finite set $V$ of cardinality $n$
and a collection $\mathcal{B}$ of subsets of $V$ such that every $l$-subset of $V$ appears in at least $\lambda$ elements of $\mathcal{B}$
and such that for any element $B \in \mathcal{B}$, the cardinality $\vert B\vert$ is in $K_\mu$.
When $K_\mu$ is the singleton $\{\mu\}$, an $l$-$(n,K_\mu,\lambda)$ generalized covering reduces to an $l$-$(n,\mu,\lambda)$ covering.
A generalized covering may also be seen as a straightforward generalization of an $l$-\textit{wise balanced design},
where each $l$-subset of $V$ occurs exactly $\lambda$ times \cite{:2007b}.

As in the standard covering number $C_{\lambda}(n,\mu,l)$,
we define the \textit{generalized covering number} $C_{\lambda}(n,K_\mu,l)$ to be the cardinality of a smallest $\mathcal{B}$ such that there exists an $l$-$(n,K_\mu,\lambda)$ generalized covering $(V, \mathcal{B})$.
We exploit the following lower bound on $C_{\lambda}(n,K_\mu,l)$.
\begin{proposition}\label{upb:coveringdesign}
For any positive integers $n$, $\mu$, $l$, $\lambda$ such that $n \geq \mu \geq l$ and any set $K_\mu$ of positive integers with $\mu$ being its largest element,
\begin{align*}
C_{\lambda}&(n,K_\mu,l)\\
&\geq \left\lceil\frac{n}{\mu}\left\lceil\frac{n-1}{\mu-1}\cdots\left\lceil\frac{n-l+2}{\mu-l+2}\left\lceil\frac{\lambda(n-l+1)}{\mu-l+1}\right\rceil\right\rceil\cdots\right\rceil\right\rceil.
\end{align*}
\end{proposition}
\begin{IEEEproof}
Let $(V, \mathcal{B})$ be an $l$-$(n,K_\mu,\lambda)$ generalized covering.
For any element $B \in \mathcal{B}$,
let $B' = B \cup C_B$, where $C_B \subset V\setminus B$ is an arbitrary subset of cardinality $\mu - \vert B\vert$.
Define $\mathcal{B}' = \{B' \mid B \in \mathcal{B}\}$.
Then, the pair $(V, \mathcal{B}')$ is an $l$-$(n,\mu,\lambda)$ covering with $\vert \mathcal{B}\vert = \vert \mathcal{B}'\vert$,
which implies that $C_{\lambda}(n,K_\mu,l) \geq C_{\lambda}(n,\mu,l)$.
Applying the Sch\"{o}nheim bound for coverings given in \cite{Schonheim:1964} proves the assertion.
\end{IEEEproof}

By noticing a simple relation between $l$-separating parity-check matrices and generalized coverings, we obtain the following lower bound on separating redundancy.
\begin{theorem}\label{thm:lwb}
Let $\mathcal{C}$ be an $[n,k,d]_q$ linear code with dual distance $d^{\perp}$.
For any $l \leq \min\{d,n-k\}-1$, the $l$-separating redundancy $s_l(\mathcal{C})$ satisfies
\begin{align*}
s_l(\mathcal{C})
\geq \left\lceil\frac{n}{\nu}\left\lceil\frac{n-1}{\nu-1}\cdots\left\lceil\frac{n-l+2}{\nu-l+2}\left\lceil\frac{\lambda(n-l+1)}{\nu-l+1}\right\rceil\right\rceil\cdots\right\rceil\right\rceil,
\end{align*}
where $\nu = n -d^{\perp}$ and $\lambda = n-k-l$.
\end{theorem}
\begin{IEEEproof}
Let $H$ be an\, $m \times n$\, $l$-separating parity-check matrix for $\mathcal{C}$ and $\boldsymbol{r}_i$ its $i$th row for $0 \leq i \leq m-1$.
Regard each row $\boldsymbol{r}_i = (r^i_0,\dots,r^i_{n-1})$, $r^i_j \in \mathbb{F}_q,$ as an $n$-dimensional vector in $\mathcal{C}^{\perp}$.
Let $\overline{\operatorname{supp}}(\boldsymbol{r}_i) = \{j \mid r^i_j = 0, 0 \leq j \leq n-1\}$ be the complement of the support of $\boldsymbol{r}_i$
and define a multiset $\mathcal{B} = \{\overline{\operatorname{supp}}(\boldsymbol{r}_i) \mid 0 \leq i \leq m-1\}$.
For a set $S\subset\{0,1,\dots,n-1\}$, define $M_S$ to be the submatrix of $H$ obtained by discarding all rows that contain a nonzero element in at least one coordinate in $S$.
Because $H$ is $l$-separating for any $S \subset \{0,1,\dots,n-1\}$ with $\vert S\vert = l$, we have $\operatorname{rank}(M_S) = n-k-l$.
Therefore, $M_S$ contains at least $n-k-l$ rows.
Note that the submatrix of $M_S$ that consists of the columns indexed by the elements of $S$ is the zero matrix.
Thus, any $S$ appears at least $n-k-l$ times as a subset of an element of $\mathcal{B}$,
which implies that the pair $(\{0,1,\dots,n-1\}, \mathcal{B})$ is an $l$-$(n,K_\mu,n-k-l)$ generalized covering where $K_{\mu}$ is a finite set of positive integers whose largest element is $\mu \leq n-d^{\perp}$.
Hence, letting
\begin{align*}
L_{\lambda}&(x,y,z)\\
&= \left\lceil\frac{x}{y}\left\lceil\frac{x-1}{y-1}\cdots\left\lceil\frac{x-z+2}{y-z+2}\left\lceil\frac{\lambda(x-z+1)}{y-z+1}\right\rceil\right\rceil\cdots\right\rceil\right\rceil
\end{align*}
for positive integers $x$, $y$, $z$, and $\lambda$ with $x \geq y \geq z$, by Proposition \ref{upb:coveringdesign}, we have
\begin{align*}
s_l(\mathcal{C}) &\geq C_{n-k-l}(n,K_\mu,l)\\
&\geq L_{n-k-l}(n,\mu,l)\\
&\geq L_{n-k-l}(n,n-d^{\perp},l),
\end{align*}
as desired.
\end{IEEEproof}

It is notable that if we omit the ceiling functions in Theorem \ref{thm:lwb} to bound the right-hand side of the inequality from below, our lower bound reduces to the volume bound in Theorem \ref{lwb:volume}.
Hence, our bound is quite often sharper than Theorem \ref{lwb:volume} and always at least as sharp.

\subsection{Probabilistic upper bounds}\label{subsec:upb}
We now turn our attention to upper bounds on the separating redundancy of a linear code.
In what follows, for a pair $x$, $y$ of nonnegative integers $x\ge y$,
\[
{x\brack y}_q=\prod_{i=0}^{y-1}\frac{q^{x-i}-1}{q^{i+1}-1}
\]
is defined to be the Gaussian binomial coefficient, which counts the number of $y$-dimensional subspaces in an $x$-dimensional subspace over $\mathbb{F}_q$.

To present our idea in a simple manner, we first prove the following basic upper bound through a probabilistic argument.
\begin{theorem}\label{thm:probabilistic}
Let $\mathcal{C}$ be an $[n,k,d]_q$ linear code.
For any $l \leq \min\{d,n-k\}-1$, the $l$-separating redundancy $s_l(\mathcal{C})$ satisfies
\[s_l(\mathcal{C}) \leq \min_{t\in\mathbb{N}}\left\{t + \left\lfloor \binom{n}{l}\sum_{r = 0}^{n-k-l}(n-k-l-r)P_{t,r}\right\rfloor\right\},\]
where
\[P_{t,r} = \sum_{i=r}^{t}\binom{t}{i}(1-q^{-l})^{t-i}\frac{{n-k-l \brack r}_q\prod_{j=0}^{r-1}(q^i-q^j)}{q^{i(n-k)}}.\]
\end{theorem}

As we will later illustrate with numerical examples, the above basic bound is already sharper for interesting linear codes than Theorem \ref{upb:probabilistic},
which is the strongest among the known ones in many cases.
Although it is not easy to directly compare Theorems \ref{thm:probabilistic} and \ref{upb:probabilistic} in general,
to explain how our probabilistic approach is related to Theorem \ref{upb:probabilistic},
we also prove an even sharper but slightly more complicated variant of Theorem \ref{thm:probabilistic},
which will be given as Theorem \ref{thm:probabilistic2} later in this section.
In addition to these, one more variant of Theorem \ref{thm:probabilistic} is given to present a slightly better upper bound for the case when the alphabet size $q$ is large.
This version will be referred to as Theorem \ref{thm:probabilisticYu}.

Now, to prove Theorem \ref{thm:probabilistic} and its two variants, we define a special kind of combinatorial matrix.
An \textit{orthogonal array} OA$(m,n,g,l)$ is an $m\times n$ matrix over a finite set $\Gamma$ of cardinality $g$ such that in any $m\times l$ submatrix every $l$-dimensional vector in $\Gamma^l$ appears exactly $\frac{m}{g^l}$ times as a row.
It is a well-known fact that an OA can by constructed by using the codewords of a linear code as rows,
which may be seen as a corollary of Delsarte's equivalence theorem \cite[Theorem 4.5]{Delsarte:1973}.
\begin{proposition}[\cite{Kempthorne:1947}]\label{prop:oa}
Let $\mathcal{C}$ be an $[n,k,d]_q$ linear code over $\mathbb{F}_q$. A $q^{n-k}\times n$ matrix formed by all codewords of $\mathcal{C}^\perp$ as rows is an \textup{OA}$(q^{n-k},n,q,d-1)$.
\end{proposition}

We employ a well-known fact on the ranks of matrices over a finite field.
For various known proofs of the following lemma, see, for example, \cite{Landsberg:1893,Fisher:1966}.
\begin{lemma}[\cite{Landsberg:1893}]\label{lm:subspace}
Take $t, u, v \in \mathbb{N}$ with $u \geq v$.
Let $M$ be a $t\times u$ matrix whose rows are drawn independently and uniformly at random from a $v$-dimensional subspace in a $u$-dimensional 
subspace over $\mathbb{F}_q$. For $0\le r\le v$, the probability that $M$ is of rank $r$ over $\mathbb{F}_q$ is 
\[\frac{{v\brack r}_q\prod_{i=0}^{r-1}(q^t-q^i)}{q^{vt}}.\]
\end{lemma}

Note that the above lemma is usually stated as an enumeration formula for $t \times v$ matrices of rank $r$ over a finite field.
In fact, because a pair of subspaces of the same finite dimension over the same finite field are isomorphic,
the $u$-dimensional ambient space in Lemma \ref{lm:subspace} is vacuous in the sense that we may as well consider $t \times v$ matrices of a given rank with rows chosen from $\mathbb{F}_q^v$.
However, because we apply a probabilistic argument to a linear code and its punctured code, we consider a subspace in a larger subspace and state the formula in probabilistic language.

To present our proof in a concise manner, we slightly generalize the concept of an $S$-separating matrix by allowing the matrix under consideration to be a parity-check matrix for a supercode.
Let $A$ be a parity-check matrix for a supercode of a given $[n,k,d]_q$ linear code $\mathcal{C}$.
Define $A(S)$ to be the submatrix of $A$ obtained by discarding all rows with a nonzero element in at least one coordinate in $S$ and deleting all columns corresponding to the coordinates in $S$.
If $S$ is the empty set, we define $A(S) = A$.
The parity-check matrix $A$ for the supercode is $S$-\textit{separating with respect to} $\mathcal{C}$ if $A(S)$ is a parity-check matrix for $\mathcal{C}_{\overline{S}}$.
It is easy to see that for $\vert S\vert \leq d-1$, the matrix $A$ is $S$-separating with respect to $\mathcal{C}$ if and only if $\operatorname{rank}(A(S)) = n-k-\vert S\vert$.
It should be noted that $A$ can be a parity-check matrix for $\mathcal{C}$ itself because technically the notion of a supercode $\mathcal{D}$ of $\mathcal{C}$ admits the case $\mathcal{D} = \mathcal{C}$.

The following simple lemma plays a key role in our probabilistic argument.
\begin{lemma}\label{lm:superPMisPM}
Let $A$ be a parity-check matrix for a supercode of an $[n,k,d]_q$ linear code $\mathcal{C}$ and take a positive integer $l \leq \min\{d,n-k\}-1$.
If $A$ is $S$-separating with respect to $\mathcal{C}$ for any $S \subset \{0,1,\dots,n-1\}$ of cardinality $l$, then $A$ is an $l$-separating parity-check matrix for $\mathcal{C}$.
\end{lemma}
\begin{IEEEproof}
Let $S' \subset \{0,1,\dots,n-1\}$ be a set of $l-1$ coordinates.
Because $\operatorname{rank}(A(S)) = n-k-l > 0$ for any $S \subset \{0,1,\dots,n-1\}$ of cardinality $l$,
the matrix $A(S')$ contains at least one nonzero entry.
Without loss of generality, we assume that the $i$th column of $A$ contains a nonzero element in a row of $A(S')$, where $i \not\in S'$.
Because $A$ is $(S'\cup\{i\})$-separating with respect to $\mathcal{C}$ by assumption, we have $\operatorname{rank}(A(S'\cup\{i\})) = n-k-l$.
Thus, since $i \not\in S'$ and the $i$th column of $A$ contains a nonzero element in a row of $A(S')$,
we have $\operatorname{rank}(A(S')) \geq n-k-l+1$.
However, because $A(S')$ is a parity-check matrix for a supercode of $\mathcal{C}_{\overline{S'}}$, we have $\operatorname{rank}(A(S')) \leq n-k-l+1$,
which implies that $\operatorname{rank}(A(S'))= n-k-l+1$.
Hence, $A$ is $S'$-separating with respect to $\mathcal{C}$.
By induction, for any $S'' \subset \{0,1,\dots,n-1\}$ with $1\le \vert S''\vert \le l$, the matrix $A$ is $S''$-separating with respect to $\mathcal{C}$.
Consider a singleton $\{j\} \subset \{0,1,\dots,n-1\}$.
Because $A$ is $\{j\}$-separating with respect to $\mathcal{C}$, we have $\operatorname{rank}(A(\{j\})) = n-k-1$.
If the $j$th column of $A$ contains a nonzero element, then $A$ is of rank $n-k$ and thus a parity-check matrix for $\mathcal{C}$.
Therefore, if there exists a column of $A$ that contains a nonzero element, then we are done.
If all columns are zero column vectors, then $A$ is the zero matrix, which contradicts the assumption that $A$ is $S$-separating with respect to $\mathcal{C}$ for any $S \subset \{0,1,\dots,n-1\}$ of cardinality $l$.
\end{IEEEproof}

We are now ready to present the probabilistic proof of our basic upper bound.
In what follows, the expected value of a given random variable $X$ is denoted by $\mathbb{E}(X)$.
\begin{IEEEproof}[Proof of Theorem \ref{thm:probabilistic}]
We employ a version of the probabilistic proof technique, which is known as the \textit{alternation method} \cite{Alon:2016}
and the \textit{sample-and-modify} technique \cite{Mitzenmacher:2017}.
Construct a $t \times n$ matrix $A$ by taking independently and uniformly at random $t$ codewords from $\mathcal{C}^{\perp}$ as rows
and let $\boldsymbol{a}_m$ be its $m$th row for $0\le m\le t-1$.
We claim that a small modification to this random matrix gives an $l$-separating parity-check matrix.
Let $\mathcal{S} = \{S \subset \{0,1,\dots,n-1\} \mid \vert S\vert = l\}$ be the set of $\binom{n}{l}$ subsets $S \subset \{0,1,\dots,n-1\}$ of cardinality $l$.
For given $S\in\mathcal{S}$, define $C_S = \{(c_0, \dots, c_{n-1})\in\mathcal{C}^{\perp} \mid c_j =0, j\in S\}$. 
By Proposition \ref{prop:oa}, the probability that $\boldsymbol{a}_m\in C_S$ is $q^{-l}$. Therefore, the probability that exactly $i$ rows of $A$ are in $C_S$ is 
\[
\binom{t}{i}q^{-li}(1-q^{-l})^{t-i}.
\]
Note that $\vert C_S \vert =q^{n-k-l}$.
Hence, by Lemma \ref{lm:subspace}, the probability $p_{A,S,r}$ that for any $S \in \mathcal{S}$, $A(S)$ is of rank $r$ is
\begin{align*}
p_{A,S,r} &= \sum_{i = r}^{t}\binom{t}{i}q^{-li}(1-q^{-l})^{t-i}\frac{{n-k-l\brack r}_q\prod_{j=0}^{r-1}(q^i-q^j)}{q^{i(n-k-l)}}\\
&= P_{t,r}.
\end{align*}
We adjoin more rows if $\operatorname{rank}(A(S))$ is less than $n-k-l$.
Let $X$ be the random variable counting the smallest number of additional rows required to attach to $A$ to turn it into an $S$-separating matrix with respect to $\mathcal{C}$ for any $S \in \mathcal{S}$.
The realization of $X$ depends on $A$, while its expectation $\mathbb{E}(X)$ is a function of $t$.
Trivially, the probability that $X \leq \mathbb{E}(X)$ is strictly positive.
Thus, by Lemma \ref{lm:superPMisPM}, with positive probability, appending appropriately chosen rows to $A$ gives an $l$-separating parity-check matrix for $\mathcal{C}$ with at most $t+\lfloor\mathbb{E}(X)\rfloor$ rows.
Therefore, we have
\begin{align}\label{ineq1}
s_l(\mathcal{C}) \leq \min_{t\in\mathbb{N}}\left\{t+\lfloor\mathbb{E}(X)\rfloor\right\}.
\end{align}
To bound the expected value $\mathbb{E}(X)$ on the right-hand side from above, notice that
\begin{align*}
\mathbb{E}(X) &\leq \mathbb{E}\left(\sum_{S \in \mathcal{S}}(n-k-l-\operatorname{rank}(A(S)))\right)\\
&= \binom{n}{l}\sum_{r=0}^{n-k-l}(n-k-l-r)P_{t,r}.
\end{align*}
Plugging in the above upper bound into Inequality (\ref{ineq1}) completes the proof.
\end{IEEEproof}

While we focused on a simple presentation of our idea, the probabilistic upper bound in Theorem \ref{thm:probabilistic} may be improved by more careful analyses.
As stated earlier, we present two useful variants that do not require too involved an argument.

Recall that Theorem \ref{upb:probabilistic} involves the function $f_q(a,b)$ which counts the number of $a \times b$ matrices over $\mathbb{F}_q$ of rank $b$ that do not contain all-zero rows.
We incorporate this knowledge to replace in our probabilistic argument the elementary fact, which is Lemma \ref{lm:subspace}, with the following lemma that allows for using a slightly more favorable probability space.
\begin{lemma}\label{lm:subspaceExcept0}
Take $t, u, v \in \mathbb{N}$ with $u \geq v$.
Let $\mathcal{E}$ be a $v$-dimensional subspace in a $u$-dimensional vector space over $\mathbb{F}_q$ and
$M$ a $t\times u$ matrix whose rows are drawn independently and uniformly at random from $\mathcal{E}\setminus\{\boldsymbol{0}\}$.
For $0\le r\le v$, the probability that $M$ is of rank $r$ over $\mathbb{F}_q$ is 
\[\frac{{v \brack r}_qf_q(t,r)}{(q^v-1)^t}.\]
\end{lemma}
\begin{IEEEproof}
For any $r$-dimensional subspace $\mathcal{F}$ of $\mathcal{E}$, take an $r\times u$ matrix $B_{\mathcal{F}}$ whose rows form a basis of $\mathcal{F}$.
Every $t \times u$ matrix $N$ of rank $r$ over $\mathbb{F}_q$ without all-zero rows can be written uniquely as a product $N = RB_{\mathcal{F}}$ of two matrices for some $\mathcal{F}$, where $R$ is a $t \times r$ matrix of rank $r$ over $\mathbb{F}_q$ without all-zero rows.
Hence, $N$ and the pair $(R, B_{\mathcal{F}})$ has a one-to-one correspondence.
Because $M$ is taken uniformly at random from all possible $(q^v-1)^t$ matrices, the claim follows by dividing the number of pairs $(R, B_{\mathcal{F}})$ by $(q^v-1)^t$.
\end{IEEEproof}

Because the zero codeword $\boldsymbol{0}$ in the dual code $\mathcal{C}^{\perp}$ does not contribute to anything when present in a parity-check matrix for $\mathcal{C}$ other than inflating the number of rows,
use of the above lemma leads to a better upper bound.
\begin{theorem}\label{thm:probabilistic2}
Let $\mathcal{C}$ be an $[n,k,d]_q$ linear code.
For any $l \leq \min\{d,n-k\}-1$, the $l$-separating redundancy $s_l(\mathcal{C})$ satisfies
\[s_l(\mathcal{C}) \leq \min_{t\in\mathbb{N}}\left\{t + \left\lfloor \binom{n}{l}\sum_{r = 0}^{n-k-l}(n-k-l-r)Q_{t,r}\right\rfloor\right\},\]
where
\[Q_{t,r} = \sum_{i=r}^{t}\binom{t}{i}c^{i}(1-c)^{t-i}\frac{{n-k-l \brack r}_qf_q(i,r)}{(q^{n-k-l}-1)^i}\]
with $c = \frac{q^{n-k-l}-1}{q^{n-k}-1}$.
\end{theorem}
\begin{IEEEproof}
Construct a $t \times n$ matrix $A$ by taking independently and uniformly at random $t$ codewords from $\mathcal{C}^{\perp}\setminus\{\boldsymbol{0}\}$ as rows.
Argue in the same manner as in the proof of Theorem \ref{thm:probabilistic} by using Lemma \ref{lm:subspaceExcept0} in place of Lemma \ref{lm:subspace}.
\end{IEEEproof}

To see how Theorem \ref{thm:probabilistic2} is related to Theorem \ref{upb:probabilistic},
let $e(t,\mathcal{C})$ and $e(t,\mathcal{C},S)$ be the number of $t \times n$ parity-check matrices with no all-zero rows for a given $[n,k,d]_q$ linear code $\mathcal{C}$ and that of $S$-separating ones with no all-zero rows for a given subset $S\subset\{0,1,\dots,n-1\}$, respectively.
The proof of Theorem \ref{upb:probabilistic} calculates $e(t,\mathcal{C})$ and $e(t,\mathcal{C},S)$, which also shows that $e(t,\mathcal{C},S) = e(t,\mathcal{C},S')$ for any pair $S$, $S'$ with $\vert S\vert = \vert S'\vert$.
Hence, for any subset $S \subset \{0,1,\dots,n-1\}$ of cardinality $l$, we may safely write $e(t,\mathcal{C},l)$ to mean the number of $S$-separating parity-check matrices for $\mathcal{C}$ with $t$ rows.
With this notation, the pigeonhole principle ensures that there exists at least one\, $t \times n$\, $l$-separating parity-check matrix for $\mathcal{C}$ if
\begin{align}\label{ineq:piegon}
e(t,\mathcal{C}) > \binom{n}{l}(e(t,\mathcal{C}) - e(t,\mathcal{C},l)).
\end{align}
Theorem \ref{upb:probabilistic} is a claim that the $l$-separating redundancy of $\mathcal{C}$ must be smaller than or equal to the smallest possible $t$ that satisfies the above inequality.

It is well known that a counting argument of this kind can be translated into a probabilistic one either through the union bound or through linearity of expectation.
For our purpose, it is more convenient to choose the latter.

Let $\mathcal{S} = \{S \subset \{0,1,\dots,n-1\} \mid \vert S\vert = l\}$ be the set of $\binom{n}{l}$ subsets $S \subset \{0,1,\dots,n-1\}$ of cardinality $l$.
Take a $t\times n$ parity-check matrix $H$ uniformly at random from the set of $e(t,\mathcal{C})$ parity-check matrices with no all-zero rows for $\mathcal{C}$.
Define $X_S$ to be the random variable that equals $0$ if $H$ is $S$-separating and $1$ otherwise.
Let $X = \sum_{S\in\mathcal{S}}X_S$. If $\mathbb{E}(X) < 1$, then there exists a\, $t\times n$\, $l$-separating parity-check matrix,
which implies that if
\begin{align*}
1 &> \mathbb{E}(X)\\
&= \sum_{S\in\mathcal{S}}\mathbb{E}(X_S)\\
&= \binom{n}{l}\left(1-\frac{e(t,\mathcal{C},l)}{e(t,\mathcal{C})}\right),
\end{align*}
then there exists an $l$-separating parity-check matrix for $\mathcal{C}$ with $t$ rows.
Clearly, the above inequality is equivalent to Inequality (\ref{ineq:piegon}),
proving the same bound by a probabilistic argument.

In the proofs of Theorems \ref{thm:probabilistic} and \ref{thm:probabilistic2},
we randomly sample a parity-check matrix $A$ for a supercode of $\mathcal{C}$.
Note that if $A$ has a large number of rows, it is very likely $l$-separating and requires virtually no additional rows.
Hence, if we took a sufficiently large $A$, then the argument would be nearly identical to the probabilistic version of the proof of Theorem \ref{upb:probabilistic} except that
with our approach there would be a very tiny probability that the chosen $A$ lacked enough linearly independent rows to be a parity-check matrix for $\mathcal{C}$.
The trick we used for our upper bounds is that we deliberately pick rather small $A$ and, if it is not $S$-separating for some $S\in\mathcal{S}$, we fix the blemishes by appending a few more rows.
Once we finish making our small matrix $S$-separating for any $S\in\mathcal{S}$, Lemma \ref{lm:superPMisPM} assures that the modified matrix is automatically a parity-check matrix for $\mathcal{C}$ rather than for its proper supercode with a larger dimension.

The crucial point is that, unlike the probabilistic proof of Theorem \ref{upb:probabilistic}, the proofs of Theorems \ref{thm:probabilistic} and \ref{thm:probabilistic2} do not require a probability space in which a randomly chosen matrix is typically $l$-separating.
Hence, our bounds can be sharper as long as we can make a good estimate of the required number of additional rows.

One weakness of Theorems \ref{thm:probabilistic} and \ref{thm:probabilistic2} is that these probabilistic bounds depend on $q$.
Indeed, they become looser as the alphabet size $q$ increases and give larger estimates for nonbinary linear codes.
This contrasts with the fact that the lower bound in Theorem \ref{thm:lwb} is independent of $q$.
In the remainder of this subsection, we show that a simple trick can give an alternative bound that mitigates this weakness to some extent.

\begin{theorem}\label{thm:probabilisticYu}
Let $\mathcal{C}$ be an $[n,k,d]_q$ linear code.
For any $l \leq \min\{d,n-k\}-1$, the $l$-separating redundancy $s_l(\mathcal{C})$ satisfies
\[s_l(\mathcal{C}) \leq n-k+\min_{t\in\mathbb{N}}\left\{t+\left\lfloor b_{l}\!\sum_{r = 0}^{n-k-l}\!(n-k-l-r)Q_{t,r}\right\rfloor\right\},\]
where $b_l = \binom{n}{l}-\binom{n-k}{l}$ and
\[Q_{t,r} = \sum_{i=r}^{t}\binom{t}{i}c^{i}(1-c)^{t-i}\frac{{n-k-l \brack r}_qf_q(i,r)}{(q^{n-k-l}-1)^i}\]
with $c = \frac{q^{n-k-l}-1}{q^{n-k}-1}$.
\end{theorem}
\begin{IEEEproof}
Take an $(n-k)\times n$ parity-check matrix $H$ for $\mathcal{C}$ in standard form,
so that $H$ contains the $(n-k)\times(n-k)$ identity matrix $I$ as its submatrix.
Let $T$ be the set of coordinates that index the columns of $I$ in $H$.
Because $I$ contains exactly one nonzero element in each column and in each row,
we have $\operatorname{rank}(H(S)) = n-k-l$ for any $l$-subset $S\subset T$.
Hence, $H$ is $S$-separating for any $S\subset T$ of cardinality $l$.
Carrying out the same argument as in the proof of Theorem \ref{thm:probabilistic2} over the remaining $\binom{n}{l}-\binom{n-k}{l}$ $l$-subsets of coordinates proves the assertion.
\end{IEEEproof}

Theorem \ref{thm:probabilisticYu} reduces the coefficient of the sum from $\binom{n}{l}$ to $\binom{n}{l}-\binom{n-k}{l}$ in exchange for the newly introduced constant additive term $n-k$.
While the additional $n-k$ rows is a nontrivial penalty, as we will illustrate with a nonbinary linear code, the benefit of the smaller coefficient outweighs the disadvantage in the constant additive term when the sum of $rQ_{r,t}$ over all $r$ is small.

\subsection{Refined design-theoretic upper bound}\label{subsec:coveringupb}
Probabilistic combinatorics provides powerful tools for proving the existence of a desired mathematical object.
However, verifying the existence alone does not necessarily supply an efficient algorithm for construction.
While our probabilistic upper bounds are general and quite strong compared to the other known general upper bounds, their proofs do not give any insight into how we may be able to efficiently construct $l$-separating parity-check matrices that achieve the bounds.

To address this disadvantage, we refine the design-theoretic upper bound given in Theorem \ref{upb:covering}.
As long as a suitable covering can be constructed efficiently, our design-theoretic bound is constructive and, in some cases, sharper than the probabilistic ones.
In the following theorem, $C_\lambda(n,\mu,l)$ is the standard covering number as we defined in Section \ref{subsec:lwb}.

\begin{theorem}\label{thm:upbcovering}
Let $\mathcal{C}$ be an $[n,k,d]_q$ linear code.
Define $\mu=\min\{d,n-k\}-1$. For $l \leq \mu$,
the $l$-separating redundancy $s_l(\mathcal{C})$ satisfies
\[s_l(\mathcal{C}) \leq \min\left\{(n-k)C_{1}(n,\mu,l), (n-k-l)\binom{n}{l}\right\}.\]
\end{theorem}
\begin{IEEEproof}
Let $(\{0,1,\dots,n-1\},\mathcal{B})$ be an $l$-$(n,\mu,1)$ covering and $H_{\text{all}}$ a $q^{n-k} \times n$ parity-check matrix for $\mathcal{C}$ whose rows are the $q^{n-k}$ codewords of $\mathcal{C}^{\perp}$.
By Proposition \ref{prop:oa}, for any $\mu$-subset $S \subset \{0,1,\dots,n-1\}$, the parity-check matrix $H_{\text{all}}$ contains a $\mu \times n$ submatrix $I_S$ whose columns indexed by the elements of $S$ form the $\mu \times \mu$ identity matrix.
Because $\operatorname{rank}(H_{\text{all}}) = n-k$, it also contains an $(n-k-\mu) \times n$ submatrix $M_S$ of rank $n-k-\mu$ over $\mathbb{F}_q$ whose columns indexed by the elements of $S$ form the zero matrix.
It is straightforward to see that the $(n-k) \times n$ matrix $H_S$ obtained by stacking $I_S$ on top of $M_S$ is an $S$-separating parity-check matrix for $\mathcal{C}$.
Taking $H_B$ for all $B\in\mathcal{B}$ and stacking them on top of each other gives an $l$-separating parity-check matrix with $(n-k)\vert\mathcal{B}\vert$ rows.
Therefore, we have $s_l(\mathcal{C}) \leq (n-k)C_1(n,\mu,l)$.
To obtain an $l$-separating parity-check matrix with $(n-k-l)\binom{n}{l}$ rows, stack $M_S$ for all $S \subset \{0,1,\dots,n-1\}$ of cardinality $l$.
\end{IEEEproof}

For the above theorem to be effective, we need a covering that achieves or almost attains the covering number.
For small $l$ and $\mu$, numerous explicit constructions for such $l$-$(n,\mu,\lambda)$ coverings are known.
Indeed, direct constructions for $l = 2$ and $\mu = 3$ have been known since the mid-19th century \cite{Kirkman:1847}.
However, for $l \geq 5$, it does not seem easy to give a direct construction for a nontrivial covering $(V, \mathcal{B})$ with the smallest possible $\vert \mathcal{B}\vert$.
Nonetheless, there are also various known algorithms that produce suboptimal coverings for $l \geq 5$ with fairly small $\vert \mathcal{B}\vert$,
with one notable example being the heuristic \textsf{TS-CD} algorithm \cite{Fadlaoui:2011}.
For complete treatments of this topic, we refer the reader to \cite[Section VI-11]{:2007b} and references therein.

To see how Theorem \ref{thm:upbcovering} improves Theorem \ref{upb:covering}, consider the following proof of the latter.
First, as in the proof of Theorem \ref{thm:upbcovering}, take an $l$-$(n,\mu,1)$ covering $(\{0,1,\dots,n-1\},\mathcal{B})$ and stack $M_S$ for $S\in\mathcal{B}$ on top of each other.
This ensures that the resulting matrix $H$ satisfies $\operatorname{rank}(H(S')) \geq n-k-\mu$ for any $l$-subsets $S' \subset \{0,1,\dots,n-1\}$.
To turn $H$ into an $l$-separating one, we may simply adjoin a $(\mu-l) \times n$ matrix $I_{S'}$ for all possible $\binom{n}{l}$ patterns of $S'$,
which establishes the same upper bound as in Theorem \ref{upb:covering}.

Note that while we can use an arbitrary covering in the first step of the above alternative proof, the second step uses the poorest, trivial covering that consists of the $\binom{n}{l}$ $l$-subsets themselves.
The key idea in the proof of Theorem \ref{thm:upbcovering} is that we can also exploit the same nontrivial covering in the second step so that we do not need to fix each of the $\binom{n}{l}$ blemishes one by one.
Indeed, use of a covering with $\vert \mathcal{B}\vert < \frac{\mu-l}{\mu}\binom{n}{l}$ always results in a smaller $l$-separating parity-check matrix.

The benefit of the refinement we described above disappears if $l = \mu = \min\{d,n-k\}-1$.
In this degenerate case, the first step alone gives an $l$-separating parity-check matrix, so that
Theorems \ref{thm:upbcovering} and \ref{upb:covering} both reduce to the same bound $s_l(\mathcal{C}) \leq (n-k-l)\binom{n}{l}$.

It is worth mentioning that combining Theorems \ref{thm:upbcovering} and \ref{upb:covering} gives a simple and sharper expression.
To see this, notice that arguing the same way as in the proof of Proposition \ref{upb:coveringdesign} shows that $C_1(n,\mu,l) \leq C_1(n,\mu',l)$ for $\mu \geq \mu'$.
It is proved in \cite{Abdel-Ghaffar:2013} that the result of the minimization on the right-hand side of the inequality in Theorem \ref{upb:covering} is strictly smaller than $(n-k-l)\binom{n}{l}$.
Hence, since Theorem \ref{thm:upbcovering} says that the separating redundancy is smaller than or equal to this number and $(n-k)C_{1}(n,\mu,l)$,
we have the following corollary.
\begin{corollary}\label{coro:upbcovering}
Let $\mathcal{C}$ be an $[n,k,d]_q$ linear code.
Define $\mu=\min\{d,n-k\}-1$. For $l \leq \mu$,
the $l$-separating redundancy $s_l(\mathcal{C})$ satisfies
\begin{align*}
s_l(\mathcal{C}) \leq \min\left\{\right.&(n-k)C_{1}(n,\mu,l),\\
&\left.(n-k-\mu)C_{1}(n,\mu,l)+\binom{n}{l}(\mu-l)\right\}.
\end{align*}
\end{corollary}

It is notable that we can also exploit any $l$-$(n,K_\mu,1)$ generalized covering $(\{0,1,\dots,n-1\},\mathcal{B})$, even if $\mu \geq d$ and regardless of whether $K_{\mu}$ is a singleton,
as long as for any $B \in \mathcal{B}$, the columns indexed by the elements of $B$ in a parity-check matrix for $\mathcal{C}$ are linearly independent.
Indeed, linear independence among the columns indexed by the elements of $B$ ensures that $H_{\text{all}}$ contains the key components $I_B$, $M_B$,
where $I_B$ is a $\vert B\vert \times n$ submatrix whose columns indexed by the elements of $B$ form the identity matrix and
$M_B$ is an $(n-k-\vert B\vert) \times n$ submatrix of rank $n-k-\vert B\vert$ over $\mathbb{F}_q$ whose columns indexed by the elements of $B$ form the zero matrix.
Hence, as in the proof of Theorem \ref{thm:upbcovering}, we can obtain an $l$-separating parity-check matrix by stacking $I_B$ and $M_B$ for all $B \in \mathcal{B}$.

To illustrate the generalized approach, we apply it to a class of linear codes from finite geometry.
For fundamental notions and basic facts in finite geometry, we refer the reader to \cite{Hirschfeld:1998}.

The \textit{affine geometry} AG$(m,q)$ of \textit{dimension} $m$ over $\mathbb{F}_q$ is a finite geometry whose \textit{points} are the vectors in $\mathbb{F}_q^m$ and $i$-\textit{flats} are the $i$-dimensional vector spaces of $\mathbb{F}_q^m$ and their cosets.
We use AG$(2,q)$ with $q$ even and consider its points and $1$-flats, that is, the \textit{affine plane} with $q=2^h$ and its $4^h$ points and $2^h(2^h+1)$ \textit{lines}.
As in other typical geometries such as Euclidean geometry, the finite geometry over a finite field also has the property that
for any set $S$ of five points in which no three points are collinear, there is a unique conic passing through $S$ \cite{Hirschfeld:1998}.
A conic specified this way by five points in which no three points are collinear is said to be \textit{irreducible}.
In what follows, we assume that the points and lines are both arbitrarily ordered.

The \textit{line-by-point incidence matrix} $H = (h_{i,j})$ is the $2^h(2^h+1) \times 4^h$ matrix over $\mathbb{F}_2$ whose rows and columns are indexed by the lines and points, respectively, such that
the entry $h_{i,j}$ of the $i$th row of the $j$th column is $1$ if the $i$th line passes through the $j$th point and $0$ otherwise.
It is known that the binary linear code defined by $H$ as a parity-check matrix has parameters $[4^h,4^h-3^h,2^h+2]_2$ (see \cite{Graham:1966} for the earliest proof and also \cite{Calkin:1999} for the same fact in our terminology).
Using a generalized covering tailored to this linear code, we construct a smaller $5$-separating parity-check matrix than is achievable by the construction in the proof of Theorem \ref{thm:upbcovering}.
\begin{theorem}\label{thm:AG}
Let $H$ be the line-by-point incidence matrix of affine geometry \textup{AG}$(2,2^h)$ with $h \geq 3$.
Define $\mathcal{C}$ to be the $[4^h,4^h-3^h,2^h+2]_2$ linear code obtained by using $H$ as a parity-check matrix.
Then, the $5$-separating redundancy of $\mathcal{C}$ satisfies
\[s_5(\mathcal{C}) \leq 3^h(2^{5h}+3\cdot 2^{4h-1}+9\cdot 2^{3h-1}-3\cdot 2^{h}).\]
\end{theorem}
\begin{IEEEproof}
It suffices to show that there is a suitable $5$-$(4^h,K_\mu,1)$ generalized covering $(V, \mathcal{B})$
such that for any $B \in \mathcal{B}$, the columns indexed by the elements of $B$ in a parity-check matrix for the $[4^h,4^h-3^h,2^h+2]_2$ linear code are linearly independent.
We construct such a design from lines and irreducible conics in AG$(2,2^h)$.
Let $\mathcal{L}$ be the set of lines in AG$(2,2^h)$.
For any pair $l_0, l_1 \in \mathcal{L}$ of parallel lines, take an arbitrary set $U_{l_0, l_1}^6$ of six parallel lines each of which intersects both $l_0$ and $l_1$.
Define $\mathcal{L}_0$ to be the set of irreducible conics in AG$(2,2^h)$.
Let $\mathcal{L}_1 = \{l_0 \cup l_1 \mid l_0, l_1 \in \mathcal{L}, \text{\ $l_0$ and $l_1$ are nonparallel}\}$
and $\mathcal{L}_2 = \{(l_0 \cup l_1)\setminus l_2 \mid \text{$l_0, l_1 \in \mathcal{L}$ are parallel}, l_2 \in U_{l_0,l_1}^6\}$.
Define $\mathcal{B} = \mathcal{L}_0\cup \mathcal{L}_1\cup \mathcal{L}_2$.
It is routine to show that $\vert\mathcal{L}_0\vert = 2^{5h}+2^{4h}+2^{3h}$, $\vert\mathcal{L}_1\vert = \binom{2^h+1}{2}2^{2h}$, and $\vert\mathcal{L}_2\vert = 6(2^h+1)\binom{2^h}{2}$.
Hence, because there is no overlap between $\mathcal{L}_0$, $\mathcal{L}_1$, and $\mathcal{L}_2$, we have $\vert\mathcal{B}\vert = 2^{5h}+3\cdot 2^{4h-1}+9\cdot 2^{3h-1}-3\cdot 2^{h}$.

We first show that every quintuple of points appears in an element of $\mathcal{B}$.
Let $S$ be a set of five points in AG$(2,2^h)$.
If the five points in $S$ all lie on a single line $l_0$, there exists another line $l_1 \not= l_0$ that passes through a point $p \not\in S$ on $l_0$.
Hence, $S$ appears in $l_0\cup l_1 \in \mathcal{L}_1$.
By the same token, if exactly four points in $S$ are collinear, considering a line $l_1$ that passes through the remaining point in $S$ and intersects the line $l_0$ that carries the four points, we have $S \subset l_0\cup l_1 \in \mathcal{L}_1$.
If no three points in $S$ are collinear, because $\vert S\vert = 5$, the set $S$ is contained in exactly one irreducible conic in $\mathcal{L}_0$.
The remaining case is when no four points in $S$ are collinear while $S$ contains three points on the same line.
We consider two subcases.

\emph{Case 1.} There is exactly one subset $\{p_0,p_1,p_2\} \subset S$ of three collinear points.
Let $l_0$ be the line that carries $\{p_0,p_1,p_2\}$.
If the line $l_1$ that carries the remaining two points in $S$ intersects $l_0$, then $S$ appears in $l_0 \cup l_1 \in \mathcal{L}_1$.
If $l_1$ is parallel to $l_0$, then $S$ appears in an element $(l_0\cup l_1) \setminus l_2$ of $\mathcal{L}_2$,
where $l_2 \in U_{l_0, l_1}^6$ is a line that contains no point in $S$.

\emph{Case 2.} There is more than one subset of three collinear points in $S$.
Because $\vert S\vert = 5$, we have exactly two subsets $S_0$, $S_1$ of three collinear points in $S$.
Let $l_0$ and $l_1$ be the lines that carry $S_0$ and $S_1$, respectively.
Because $l_0$ and $l_1$ are nonparallel, $S$ appears in $l_0\cup l_1 \in \mathcal{L}_1$.

It now suffices to prove that for any $B\in\mathcal{B}$, the columns of $H$ indexed by the elements of $B$ are linearly independent.
We show that every $B \in \mathcal{B}$ has a tangent at any point in $B$.
If $B \in \mathcal{L}_0$, it is an irreducible conic and hence has a tangent at any point in $B$ as desired.
If $B = l_0 \cup l_1 \in \mathcal{L}_1$ for nonparallel lines $l_0$, $l_1$, both $l_0$ and $l_1$ are tangents at their intersection $p$,
while any point $p' \not= p$ on $l_0$ or $l_1$ lies on a line parallel to $l_1$ or $l_0$, respectively, which is a tangent at $p'$ on $B$.
Finally, if $B = (l_0 \cup l_1)\setminus l_2$ for a pair $l_0$, $l_1$ of parallel lines and $l_2 \in U_{l_0,l_1}^6$,
the line that passes through a point $p$ on $l_0 \setminus l_2$ and the intersection of $l_1$ with $l_2$ is a tangent at $p$.
By symmetry, the line that passes through a point $p'$ on $l_1 \setminus l_2$ and the intersection of $l_0$ with $l_2$ is a tangent at $p'$.
\end{IEEEproof}

Considering how small a covering can be, it is straightforward to see that for $l \leq d-2$, the right-hand side of the bound in Theorem \ref{thm:upbcovering} is lower bounded by
\[\min_{\mu \in L}\left\{(n-k)C_{1}(n,\mu,l), (n-k-l)\binom{n}{l}\right\} \geq (n-k)\frac{\binom{n}{l}}{\binom{d-1}{l}}.\]
For the linear code of length $n = 4^h$ and minimum distance $d = 2^h+2$, we have
\begin{align*}
\frac{\binom{n}{5}}{\binom{d-1}{5}} &= \frac{2^h(2^h+2)(4^h-2)(4^h-3)}{2^h-3}\\
&> 2^{5h}+3\cdot 2^{4h-1}+9\cdot 2^{3h-1}-3\cdot 2^{h}.
\end{align*}
This shows that Theorem \ref{thm:AG} provides a smaller $5$-separating parity-check matrix due to the use of a more efficient generalized covering than the standard ones considered in Theorem \ref{thm:upbcovering}.

It should be noted that this approach is effective only if we can find a very small $l$-$(n,K_\mu,\lambda)$ generalized covering due to large $\mu$ compared to $l$.
In general, it is not a trivial task to find such a covering. Indeed, it seems very difficult to prove a better bound than Theorem \ref{thm:upbcovering} for the geometric code for $l \geq 6$.

\section{Numerical examples}\label{sec:tables}
In this section, we illustrate how our bounds compare against the known ones in specific cases by numerically bounding the separating redundancies of some short linear codes.
This complements the technical details on the refinements explained from the general viewpoint in the previous section.

\begin{table*}
\renewcommand{\arraystretch}{1.3}
\caption{\vspace{1mm}Bounds on the $l$-separating redundancy of the $[24,12,8]_2$ extended binary Golay code}
\label{tb1}
\centering
\begin{tabular}{ccccccccc}
\hline\hline
{\bfseries Bound} & {\bfseries Type} & $l=1$ & $l=2$ & $l=3$ & $l=4$ & $l=5$ & $l=6$ & $l=7$\\
\hline
Theorem \ref{thm:lwb} & lower & $\textbf{17}$ & $\textbf{24}$ & $\textbf{35}$ & $\textbf{50}$ & $\textbf{75}$ & $\textbf{114}$ & $\textbf{162}$\\
Theorem \ref{lwb:volume} \cite{Abdel-Ghaffar:2013} & lower & $\textbf{17}$ & $23$ & $33$ & $47$ & $69$ & $101$ & $152$\\
\hline
Theorem \ref{thm:probabilistic} & upper & $\textbf{35}$ & $\textbf{84}$ & $\textbf{185}$ & $\textbf{386}$ & $781$& $\textbf{1539}$& $2970$\\
Theorem \ref{thm:probabilistic2} & upper & $\textbf{35}$ & $\textbf{84}$ & $\textbf{185}$ & $\textbf{386}$ & $\textbf{780}$& $\textbf{1539}$& $\textbf{2969}$\\
Theorem \ref{thm:probabilisticYu} & upper & $44$ & $94$ & $195$ & $397$ & $791$& $1550$& $2980$\\
Corollary \ref{coro:upbcovering}\rlap{\textsuperscript{a}} & upper & $48$ & $204$ & $936$ & --- & --- & --- & --- \\
Theorem \ref{upb:probabilistic} \cite{Abdel-Ghaffar:2013}\rlap{\textsuperscript{a b}} & upper & $37$ & $93$ & $214$ & $466$ & $984$ & $2034$ & --- \\
Theorem \ref{upb:generic} \cite{Abdel-Ghaffar:2013} & upper & $78$ & $298$ & $793$ & $1585$ & $2509$& $3301$& $3796$\\
Theorem \ref{upb:covering} \cite{Abdel-Ghaffar:2013}\rlap{\textsuperscript{a}} & upper & $120$ & $936$ & --- & --- & --- & --- & --- \\
 \hline
 \hline
\multicolumn{9}{l}{\scriptsize\textsuperscript{a}
The cases when these upper bounds become weaker than the trivial one $s_l(\mathcal{C}) < q^{n-k}$ for a linear}\vspace{-1.1mm}\\
\multicolumn{9}{l}{\scriptsize\phantom{\textsuperscript{a}}
code $\mathcal{C}$ of length $n$ and dimension $k$ over $\mathbb{F}_q$ are marked by ``---''.}\vspace{-1.1mm}\\
\multicolumn{9}{l}{\scriptsize\textsuperscript{b}
Errors in \cite[Table I]{Abdel-Ghaffar:2013} are corrected.}\vspace{1.2mm}\\
\end{tabular}
\end{table*}
\begin{table*}
\renewcommand{\arraystretch}{1.3}
\caption{\vspace{1mm}Bounds on the $l$-separating redundancy of the $[41,33,5]_3$ ternary cyclic code}
\label{tb2}
\centering
\begin{tabular}{cccccc}
\hline\hline
{\bfseries Bound} & {\bfseries Type} & $l=1$ & $l=2$ & $l=3$ & $l=4$\\
\hline
Theorem \ref{thm:lwb} & lower & $\textbf{16}$ & $\textbf{31}$ & $\textbf{59}$ & $\textbf{113}$\\
Theorem \ref{lwb:volume} \cite{Abdel-Ghaffar:2013} & lower & $\textbf{16}$ & $29$ & $56$ & $105$\\
\hline
Theorem \ref{thm:probabilistic} & upper & $\textbf{37}$ & $\textbf{137}$ & $\textbf{445}$ & $\textbf{1366}$\\
Theorem \ref{thm:probabilistic2} & upper & $\textbf{37}$ & $\textbf{137}$ & $\textbf{445}$ & $\text{1366}$\\
Theorem \ref{thm:probabilisticYu} & upper & $44$ & $144$ & $452$ & $1374$\\
Corollary \ref{coro:upbcovering}\rlap{\textsuperscript{a}} & upper & $48$ & $1152$ & --- & ---\\
Theorem \ref{upb:probabilistic} \cite{Abdel-Ghaffar:2013} & upper & $40$ & $160$ & $558$ & $1836$\\
Theorem \ref{upb:generic} \cite{Abdel-Ghaffar:2013} & upper & $64$ & $288$ & $848$ & $1744$\\
Theorem \ref{upb:covering} \cite{Abdel-Ghaffar:2013}\rlap{\textsuperscript{b}} & upper & $113$ & $2190$ & --- & --- \\
 \hline
 \hline
\multicolumn{6}{l}{\scriptsize\textsuperscript{a}
The cases when these upper bounds become weaker than the trivial}\vspace{-1.1mm}\\
\multicolumn{6}{l}{\scriptsize\phantom{\textsuperscript{a}}
one $s_l(\mathcal{C})\! < q^{n-k}$ for a linear code $\mathcal{C}$ of length $n$ and dimension}\vspace{-1.1mm}\\
\multicolumn{6}{l}{\scriptsize\phantom{\textsuperscript{a}}
$k$ over $\mathbb{F}_q$ are marked by ``---''.}\vspace{2.3mm}\\
\end{tabular}
\end{table*}
\begin{table*}
\renewcommand{\arraystretch}{1.3}
\caption{\vspace{1mm}Bounds on the $l$-separating redundancy of the $[12,6,6]_4$ quaternary quadratic residue code}
\label{tb3}
\centering
\begin{tabular}{ccccccc}
\hline\hline
{\bfseries Bound} & {\bfseries Type} & $l=1$ & $l=2$ & $l=3$ & $l=4$ & $l=5$\\
\hline
Theorem \ref{thm:lwb} & lower & {\bfseries 10} & $\textbf{18}$ & $\textbf{36}$ & $\textbf{66}$ & $\textbf{132}$\\
Theorem \ref{lwb:volume} \cite{Abdel-Ghaffar:2013} & lower & $\textbf{10}$ & $\textbf{18}$ & $33$ & $\textbf{66}$ & $\textbf{132}$\\
\hline
Theorem \ref{thm:probabilistic} & upper & $\textbf{29}$ & $112$ & $351$ & $823$ & $\textbf{792}$\\
Theorem \ref{thm:probabilistic2} & upper & $\textbf{29}$ & $112$ & $351$ & $822$ & $\textbf{792}$\\
Theorem \ref{thm:probabilisticYu} & upper & $30$ & $111$ & $346$ & $815$ & $\textbf{792}$\\
Corollary \ref{coro:upbcovering} & upper & $30$ & $\textbf{54}$ & $\textbf{174}$ & $\textbf{608}$ & $\textbf{792}$\\
Theorem \ref{upb:probabilistic} \cite{Abdel-Ghaffar:2013}\rlap{\textsuperscript{a}} & upper & $34$ & $166$ & $688$ & $2622$ & ---\\
Theorem \ref{upb:generic} \cite{Abdel-Ghaffar:2013} & upper & $51$ & $231$ & $636$ & $1122$ & $1365$\\
Theorem \ref{upb:covering} \cite{Abdel-Ghaffar:2013} & upper & $48$ & $138$ & $334$ & $\textbf{608}$ & $\textbf{792}$\\
 \hline
 \hline
\multicolumn{7}{l}{\scriptsize\textsuperscript{a}
When $l=5$, Theorem \ref{upb:probabilistic} is weaker than the trivial bound  $s_l(\mathcal{C}) < q^{n-k}$ for}\vspace{-1.1mm}\\
\multicolumn{7}{l}{\scriptsize\phantom{\textsuperscript{a}}
a linear code $\mathcal{C}$ of length $n$ and dimension $k$ over $\mathbb{F}_q$ and is marked by ``---''.}\vspace{0.1mm}\\
\end{tabular}
\end{table*}
Tables \ref{tb1}, \ref{tb2}, and \ref{tb3} list numerical results on the lower and general upper bounds for the $[24,12,8]_2$ extended binary Golay code, $[41,33,5]_3$ ternary cyclic code with generator polynomial $g(x) = x^{30} + x^{13} + x^{2} + x + 1$, and $[12,6,6]_4$ quaternary quadratic residue code, respectively.
The best lower bound and the best upper bound for each number $l$ of erasures are highlighted in bold face.
As expected from the theoretical analyses in the previous section, our theorems consistently provide strong bounds and quite often improve the sharpest known results.
In particular, our basic probabilistic upper bound by Theorem \ref{thm:probabilistic} is already consistently strong,
while Theorems \ref{thm:probabilistic2} and \ref{thm:probabilisticYu} improve this result even further in some cases.
Our lower bound by Theorem \ref{thm:lwb} also provides a solid improvement in many cases while being always at least as tight as the strongest known one in the literature.

It is worth mentioning that while Theorem \ref{thm:probabilistic2} is always slightly better than or at least as strong as Theorem \ref{thm:probabilistic},
due to the complexity of the minimization problem involving the rather complicated function $f_q(a,b)$,
it requires significantly more computational power to numerically derive an estimate of separating redundancy.
Hence, Theorem \ref{thm:probabilistic} is useful not only for understanding the basic idea of the mathematical techniques we employed
but also for quickly computing a sharp upper bound.

It is also notable that, as can be seen in Table \ref{tb3}, the design-theoretic approach taken in Corollary \ref{coro:upbcovering} and Theorem \ref{upb:covering} can occasionally result in upper bounds that surpass all other known ones,
which is a fact that does not seem to have been emphasized in the literature.

For completeness, we note that the specialized bound given in \cite[Corollary 6]{Abdel-Ghaffar:2013} for the $1$-separating redundancy $s_1(\mathcal{C})$ of a linear code $\mathcal{C}$ of length $n$ whose dual code $\mathcal{C}^{\perp}$ has a codeword of weight $n$ gives $s_1(\mathcal{C}) \leq 22 $ for the extended Golay code,
$s_1(\mathcal{C}) \leq 21$ for the ternary cyclic code, and $s_1(\mathcal{C}) \leq 20$ for the quadratic residue code.
Except for these cases, no known specialized bounds improve or match the best results in the tables by our general theorems and corollary.

\section{Relation to X-codes}\label{sec:Xcodes}
In the remainder of this paper, we investigate a relation between parity-check matrices for error-erasure separation and matrices for efficient circuit testing.
Matrices of the latter kind are called \textit{X-codes} when discussed in the coding theory literature \cite{Lumetta:2003a,Fujiwara:2010},
which is the terminology we follow throughout this paper.
Note that these codes are not related to the special MDS array codes introduced in \cite{Xu:1999} and also happen to be called X-codes.
In what follows, $\log$ and $\ln$ will denote the binary logarithm $\log_2$ and natural logarithm $\log_e$, respectively.

This section is divided into two subsections.
In Section \ref{subsec:AAA}, we give a brief review of X-codes and explain how they are related to parity-check matrices that separate erasures from errors.
This relation is summarized in Proposition \ref{prop:xcodeseparation}.
We then prove an exponentially improved bound for X-codes in Section \ref{sec:main}.

\subsection{Background of X-codes and their relation to error-erasure separation}\label{subsec:AAA}
The original motivation of X-codes comes from integrated circuit (IC) testing,
where the primary objective is to check whether the IC under test was correctly manufactured without any faults and is working properly according to the design intention.
In typical digital circuit testing, the tester applies test patterns to the circuit under test, monitors its responses,
and declares the circuit chip defective if its output is different from what it should be.
Usually, the correct behavior is calculated beforehand by fault-free simulation of the circuit's behavior.
Simply put, this type of basic testing aims to detect a discrepancy between the observed and expected responses of the circuit under test by comparison.

Although this type of testing only requires simple comparison,
a critical problem lies in the volume of input and output data for testing a modern IC.
Indeed, the growing test-data volume required for testing a modern IC is the main driver of the cost hike due to much longer test time and large tester-memory requirements \cite{McCluskey:2003}.
X-codes are special linear functions used in a cost reduction technique, called \textit{X-compact} \cite{Mitra:2004,Mitra:2005},
where the response data from circuit under test is cleverly compressed.
In this context, a $(t,n,d,x)$ X-code hashes the $n$-bit output from the circuit under test into $t$ bits
while allowing for detecting the existence of up to $d$-bit-wise discrepancies between the actual output and correct responses
even if up to $x$ bits of the correct behavior are unknowable to the tester.

The parameter $t$ of a $(t,n,d,x)$ X-code is the \textit{length}, which represents the size of shrunk data.
The parameter $n$ corresponds to the number of bits in raw response data to be compressed at a time.
As will become clear from the mathematical definition of an X-code given later in this section,
$n$ is the number of codewords when viewed as a combinatorial binary code.
The other two parameters $d$, $x$ are measures of the guaranteed test quality by the X-code.
Hence, in general, we are interested in X-codes of shorter length $t$ with a larger number $n$ of codewords for given $d$ and $x$,
that is, codes of higher \textit{compaction ratio} $\frac{n}{t}$ or higher \textit{rate} $\frac{\log n}{t}$.

For any pair $\boldsymbol{v} = (v_0,\dots,v_{t-1}), \boldsymbol{w} = (w_0,\dots,w_{t-1}) \in \mathbb{F}_2^t$
of $t$-dimensional binary vectors over the finite field $\mathbb{F}_2$ of order $2$,
the \textit{superimposed sum} $\boldsymbol{v} \vee \boldsymbol{w}$, also known as the \textit{Boolean sum}, is the bit-by-bit OR operation that is defined to be
$\boldsymbol{v} \vee \boldsymbol{w} = (v_0 \vee w_0, \cdots ,v_{t-1} \vee w_{t-1})$,
where $v_i \vee w_i = 0$ if  $v_i = w_i = 0$, and $1$ otherwise.
The vector $\boldsymbol{v}$ \textit{covers} the other vector $\boldsymbol{w}$ if $\boldsymbol{v} \vee \boldsymbol{w} = \boldsymbol{v}$.
The \textit{addition} $\boldsymbol{v} + \boldsymbol{w}$ between two vectors $\boldsymbol{v}, \boldsymbol{w} \in \mathbb{F}_2^t$
is always assumed to be the bit-by-bit addition over $\mathbb{F}_2$ as usual.

Let $t$, $n$, and $d$ be positive integers and $x$ a nonnegative integer.
A $(t,n,d,x)$ {\it X-code} $\mathcal{C}= \{\boldsymbol{c}_0, \dots, \boldsymbol{c}_{n-1}\}$
is a set of $n$ $t$-dimensional vectors over ${\mathbb F}_2$ such that
\begin{align*}
\left(\bigvee_{\boldsymbol{c}\in K}\boldsymbol{c}\right) \vee \left(\sum_{\boldsymbol{c}'\in J}\boldsymbol{c}'\right) \not= \bigvee_{\boldsymbol{c}\in K}\boldsymbol{c}
\end{align*}
for any pair of disjoint subsets $K$ and $J$ of $\mathcal{C}$ with $\vert K\vert=x$ and $1 \leq \vert J\vert \leq d$,
where $\bigvee_{\boldsymbol{c}\in K}\boldsymbol{c} = \boldsymbol{0}$ if $K = \emptyset$
and $\bigvee_{\boldsymbol{c}\in K}\boldsymbol{c} = \boldsymbol{c}$ if $K$ is a singleton $\{\boldsymbol{c}\}$.
The $t$-dimensional vectors $\boldsymbol{c}_i$ in $\mathcal{C}$ are the \textit{codewords} of the X-code.
In short, a $(t,n,d,x)$ X-code is a set of $n$ codewords such that for every positive integer $d' \leq d$ no superimposed sum of
any $x$ codewords covers the addition of any $d'$ codewords chosen from the rest of the $n-x$ codewords.
When we speak of a $(t,n,d,x)$ X-code, we always assume that $n \geq d+x$ to avoid the degenerate case.

A $(t,n,d,x)$ X-code with $d \geq 2$ is a $(t,n,d-1,x)$ X-code by definition.
Similarly, a $(t,n,d,x)$ X-code for $x \geq 1$ forms a $(t,n,d,x-1)$ X-code,
while a $(t,n,d,x)$ X-code with $d \geq 2$ and $x \geq 1$ is a $(t,n,d+1,x-1)$ X-code \cite{Lumetta:2003}.
Note that when $x=0$, for any subset $J$ of a $(t,n,d,0)$ X-code $\mathcal{C}$ with $1\leq \vert J\vert \leq d$,
we have $\sum_{\boldsymbol{c}'\in J}\boldsymbol{c}'\neq\boldsymbol{0}$,
which implies a $(t,n,d,x)$ X-code with $d, x\geq 0$ does not contain an all-zero codeword. Hence, an X-code cannot be a linear code.

As mentioned in \cite{Fujiwara:2010}, when $d=1$, $(t,n,1,x)$ X-codes are equivalent to well-known combinatorial structures.
For example, the definition of a $(t,n,1,x)$ X-code coincides with that of an $x$-superimposed code of length $t$ with $n$ codewords,
which is also equivalent to disjunct matrices in group testing and cover-free families in combinatorics.
For this relation of X-codes to these combinatorial objects and the known results imported from the literature in the respective fields, we refer the reader to \cite{Fujiwara:2010}.

To see the role of a $(t,n,d,x)$ X-code $\mathcal{C} = \{\boldsymbol{c}_0,\dots,\boldsymbol{c}_{n-1}\}$ as a linear function for compaction,
it is convenient to regard $\mathcal{C}$ as the $t \times n$ matrix $M = (m_{i,j})$ over $\mathbb{F}_2$ obtained by viewing each codeword as a column of $M$,
so that the entry $m_{i,j}$ of the $i$th row of the $j$th column is $1$ if the $i$th coordinate of $\boldsymbol{c}_j$ is $1$, and $0$ otherwise.
It is straightforward to see that the definition of an X-code dictates
that the corresponding binary matrix $M$ form a parity-check matrix for a linear code of length $n$ and minimum distance at least $d+1$.
Therefore, given an $n$-dimensional vector $\boldsymbol{v} \in \mathbb{F}_2^n$, which represents the expected response to a test pattern,
and another distinct $n$-dimensional vector $\boldsymbol{w} \in \mathbb{F}_2^n$ with $1 \leq \operatorname{wt}(\boldsymbol{v}+\boldsymbol{w}) \leq d$,
which represents the actual, incorrect response of the circuit under test with up to $d$ erroneous output bits,
their $t$-bit syndromes $\boldsymbol{s}_v = M\boldsymbol{v}^T$ and $\boldsymbol{s}_w = M\boldsymbol{w}^T$ are always distinct,
allowing for detecting the faulty behavior by comparing the hashes.

The last parameter $x$ is to represent how well an X-code handles the possible existence of unknowable bits in the expected response $\boldsymbol{v}$.
Such unpredictable bits can occur in modern very large integrated circuits even though the tester is often the manufacturer of the circuit under test.
When the tester cannot predict the correct value of a particular bit, it is marked as X to indicate that the bit has an \textit{unknown logic value}.
With X representing no knowledge, computation involving X is defined by $a+\mbox{X}=\mbox{X}+a=\mbox{X}$ for $a \in {\mathbb F}_2$,
$0\cdot \mbox{X}=\mbox{X}\cdot 0=0$, and $1 \cdot \mbox{X}=\mbox{X}\cdot 1 =\mbox{X}$.
Because of this arithmetic in $\mathbb{F}_2\cup\{X\}$,
even a single unknown logic value X in the original vector $\boldsymbol{v}$ can easily propagate to multiple bits in the hash $\boldsymbol{s}_v = M\boldsymbol{v}^T$,
potentially masking discrepancies between the expected and observed responses.
However, because no superimposed sum of
any $x$ codewords of a $(t,n,d,x)$ X-code covers the addition of any other $d'$ codewords for any positive integer $d' \leq d$,
the corresponding matrix $M$ ensures that even if unknown logic values render some output bits useless,
a mismatch appears between the hashes of the expected response with up to $x$ X's and the actual, faulty response with up to $d$ erroneous output bits.
For more details on X-codes and their application to response data compaction, we refer the reader to \cite{Mitra:2005,Fujiwara:2010} and references therein.

Here, we draw attention to a striking similarity between $l$-separating parity-check matrices and X-codes in the matrix view.
It is straightforward to see that a $(t,n,d,x)$ X-code is equivalent to a $t\times n$ binary matrix in which the superimposed sum of any $x$ columns does not cover the addition of any other $d$ or fewer columns.
First, notice that by defintion a $(t,n,d,x)$ X-code $H$ has no stopping sets of size at most $x+1$.
Let $E'$ be an $x'$-set of columns of $H$ where $x'\le x$ and $\boldsymbol{c}$ be a column of $H$ such that $\boldsymbol{c}\not\in E'$.
Since the superimposed sum of any up to $x$ columns of $H$ does not cover any other column of $H$, there must be a row in which $\boldsymbol{c}$ has one and the others in $E'$ are all zeros, which implies $E'\cup\{\boldsymbol{c}\}$ is not a stopping set if $H$ is seen as a parity-check matrix for a binary linear code.

Now, recall that a parity-check matrix is $x$-separating if, and only if, for any pattern of $x$ or fewer erasures, a valid parity-check matrix for the corresponding punctured code can be obtained by taking the rows that do not check any of the erased symbols and then discarding the zeros at the erased positions.
If we regard a $(t,n,d,x)$ X-code $H$ as a parity-check matrix for some binary linear code $\mathcal{C}$,
the length and minimum distance of $\mathcal{C}$ is $n$ and at least $d+x+1$, respectively.
Indeed, because a $(t,n,d,x)$ X-code is also a $(t,n,d+x,0)$ X-code,
for any subset $J$ of the set of columns of $H$ with $1\leq \vert J\vert \leq d+x$,
we have
\[
\sum_{\boldsymbol{c}\in J}\boldsymbol{c}\neq\boldsymbol{0}.
\]
Now, it is straightforward to see that the definition of X-codes dictates that for any pattern of $i$ erasures with $i \leq x$,
taking the rows that do not check any of the erased symbols and then discarding the zeros at the erased positions gives a valid parity-check matrix
for some supercode $\mathcal{D}$ of the corresponding punctured code such that the minimum distance of $\mathcal{D}$ is at least $d+x+1-i$.
Note that, in the worst case scenario, the minimum distance of the correct punctured code for the original idea of error-erasure separation is also $d+x+1-i$.
This means that, roughly speaking, we can employ an X-code for error-erasure separation in the same way as we do with an $l$-separating parity-check matrix
except that we correct errors not with the correct punctured code but with a code which is as good in terms of minimum distance.
Therefore, as summarized in the following proposition, a $(t,n,d,x)$ X-code is, in a sense, \textit{distance-wise} $x$-\textit{separating}.
In the following proposition, as introduced in Section \ref{sec:pre}, $H(S)$ for a given coordinate set $S$ is the submatrix of $H$ obtained by discarding all rows that contain a nonzero element in at least one coordinate in $S$ and deleting all columns corresponding to the coordinates in $S$.

\begin{proposition}\label{prop:xcodeseparation}
A $(t,n,d,x)$ X-code $H$ has no stopping sets of size at most $x$. Moreover, for any set $S \subseteq \{0,1,\dots,n-1\}$ of coordinate positions with $\vert S\vert \leq x$,
the submatrix $H(S)$ forms a parity-check matrix for a linear code of of length $n-\vert S\vert$ and minimum distance at least $d+x+1 - \vert S\vert$.
\end{proposition}

As is the case with $l$-separating parity-check matrices, as a linear function for data compaction, it is desirable for a $(t,n,d,x)$ X-code to have as small $t$ as possible for given $n$, $d$, and $x$.
Indeed, it is a fundamental problem in the theory of X-compact to design an X-code with the largest possible compaction ratio $\frac{n}{t}$ for specified $n$, $d$, and $x$.
As is shown in the language of probabilistic methods in \cite{Fujiwara:2010},
a simple counting argument gives the following sufficient condition for the existence of a $(t,n,d,x)$ X-code.
\begin{theorem}[{\cite[Theorem 4.6]{Fujiwara:2010}}]\label{preciseY2010}
For any positive integers $t$, $n$, and $d$ and nonnegative integer $x$ that satisfy
\begin{align*}
t\geq\frac{-\log\left(\sum_{i=1}^{d}\binom{n}{x}\binom{n-x}{i}\right)}{\log (1-2^{-x-1})},
\end{align*}
there exists a $(t,n,d,x)$ X-code.
\end{theorem}

The following is a slightly weaker but more convenient form of the above theorem.
\begin{theorem}[{\cite[Theorem 4.6]{Fujiwara:2010}}]\label{Y2010}
Let $n$ and $d$ be positive integers and $x$ a nonnegative integer such that $n \geq 2d+x$.
There exists a $(t,n,d,x)$ X-code for any
\begin{align*}
t\ge2^{x+1}(d+x)(\ln 2)\log n.
\end{align*}
\end{theorem}

As far as the authors are aware, no tighter bounds of this kind that work for any $d$ and $x$
on the length of the shortest possible $(t,n,d,x)$ X-codes can be found in the literature.

An interesting consequence of Theorem \ref{Y2010} is that an arbitrarily large compaction ratio can be achieved for any $d$ and $x$ if there is no restriction on $n$.
Hence, from a coding theoretic viewpoint, it is more convenient to consider the rate $\frac{\log n}{t}$ to capture the asymptotic behavior of X-codes.

Let $A(t,d,x)$ be the maximum number $n$ of codewords for which there exists a $(t,n,d,x)$ X-code.
Define the \textit{asymptotic optimal rate} $R(d,x)$ of X-codes for given $d$ and $x$ to be
\begin{align*}
R(d,x) = \overline{\lim_{t\rightarrow \infty}}\frac{\log A(t,d,x)}{t}.
\end{align*}

Theorem \ref{Y2010} immediately proves that
for any integers $d \geq 1$ and $x \geq 0$, we have
\begin{align*}
R(d,x) \geq \frac{1}{2^{x+1}(d+x)\ln 2}.
\end{align*}

To bound $R(d,x)$ from above, we may exploit known results on superimposed codes.
Indeed, because a $(t,n,d,x)$ X-code is a $(t,n,1,x)$ X-code by definition, any upper bound on the rate of an $x$-superimposed code serves as one for X-codes.
The sharpest known general upper bound for $x$-superimposed codes is given in \cite{Dyachkov:1982}, which dictates that
for any positive integer $x$, the asymptotic optimal rate $R(1,x)$ satisfy
\begin{align*}
R(1,x) \leq \frac{2(\ln (x+1) - \ln 2 +1)}{x^2\ln 2}.
\end{align*}

Thus, we have the following bounds for X-codes.
\begin{theorem}\label{ratebound}
For any positive integers $d$ and $x$, the asymptotic optimal rate $R(d,x)$ of X-codes satisfies
\begin{align*}
\frac{1}{2^{x+1}(d+x)\ln 2} \leq R(d,x) \leq \frac{2(\ln (x+1) - \ln 2 +1)}{x^2\ln 2}.
\end{align*}
\end{theorem}

As can be seen in the above theorem, the gap between the best upper and lower bounds on the asymptotic optimal rate is quite large.
As may be hinted by Proposition \ref{prop:xcodeseparation}, we show that the probabilistic proof technique used in Section \ref{subsec:upb} works very well to improve the known bounds.

\subsection{Bound by probabilistic alterations}\label{sec:main}
We prove an upper bound on the shortest length $t$ for a $(t,n,d,x)$ X-code by essentially the same technique as in the proof of Theorem \ref{thm:probabilistic}.
The only key difference is that we sample a random matrix rather than a random dual.
The rest of the proof is nearly identical except that we now only need to ensure the minimum distance property of each relevant submatrix rather than its rank.

\begin{theorem}\label{Th:alt}
Let $n$ and $d$ be positive integers and $x$ a nonnegative integer.
There exists a $(t, n-a, d, x)$ X-code with
\begin{align*}
a=\min_{p\in[0,1]}\left\{\left\lfloor\sum_{i=1}^{d}\binom{n}{x}\right.\right.&\binom{n-x}{i}\\
&\left.\left.\left(1-\frac{1-(1-2p)^{i}}{2}(1-p)^x\right)^t\right\rfloor\right\}.\\
\end{align*}
\end{theorem}

While Theorem \ref{Th:alt} is not an explicit bound in itself, the following slightly weaker bound in closed form can be derived from the above theorem.
\begin{theorem}\label{Co:t-lower}
Let $n$ and $d$ be positive integers and $x$ a nonnegative integer.
There exists a $(t,n,d,x)$ X-code for any
\begin{align*}
t\ge (x+1)(d+x-1)(e\ln2)\log n + 4e(\ln16-\ln3).
\end{align*}
\end{theorem}

Note that Theorem \ref{Co:t-lower} improves the coefficient $2^{x+1}(d+x)\ln 2$ of the binary logarithmic term in Theorem \ref{Y2010} to $(x+1)(d+x-1)e\ln2$.
Thus, by dividing both sides of the inequality in Theorem \ref{Co:t-lower} by $t$ and considering the limit as $t$ approaches infinity,
we obtain a lower bound on the asymptotic optimal rate that exponentially improves the one in Theorem \ref{ratebound} as follows.
\begin{theorem}\label{thm:ratelower}
For any positive integer $d$ and nonnegative integer $x$, the asymptotic optimal rate $R(d,x)$ of X-codes satisfies
\begin{align*}
\frac{1}{(x\!+\!1)(d\!+\!x\!-\!1)e\ln2}\leq R(d,x)\leq \frac{2(\ln (x\!+\!1)\!-\ln 2\!+\!1)}{x^2\ln 2}.
\end{align*}
\end{theorem}

To prove Theorems \ref{Th:alt} and \ref{Co:t-lower}, we employ the following well-known fact.
\begin{proposition}\label{Gallager}
For any positive integer $s$ and $\rho \in [0,1]$, it holds that
\begin{align*}
\sum_{i\in T}\binom{s}{i}\rho^i(1-\rho)^{s-i}=\frac{1-(1-2\rho)^{s}}{2},
\end{align*}
where $T$ is the set of positive odd integers not larger than $s$.
\end{proposition}

For the proof of the proposition, see, for example, \cite{Gallager:1963}.

We now prove Theorems \ref{Th:alt} and \ref{Co:t-lower}.

\begin{IEEEproof}[Proof of Theorem \ref{Th:alt}]
Let $V$ be the set $\{1,2, \dots , n\}$ of positive integers less than or equal to $n$ and $\mathcal{J}$ the set of subsets $J \subset V$ with $1\le \vert J\vert \le d$. 
For any $J\in\mathcal{J}$, define $\mathcal{K}_J$ to be the set of $x$-subsets of $V\setminus J$. 
Let $M = (m_{i,j})$ be a $t\times n$ random matrix over $\mathbb{F}_2$
in which each entry $m_{i,j}$ is defined to be $1$ with probability $p\in[0,1]$, and $0$ with probability $1-p$ uniformly and independently at random. 
Let $\boldsymbol{c}_{0}, \dots , \boldsymbol{c}_{n-1}$ be the columns of $M$.
For given $J\in\mathcal{J}$ and given $K\in\mathcal{K}_{J}$, define $A_{J,K}$ to be the event that $\bigvee_{k\in K}\boldsymbol{c}_k$ covers $\sum_{l\in J}\boldsymbol{c}_l$.
We consider two random variables $Y_{J,K}$ and $Y$, where
\begin{align*}
Y_{J, K} = \begin{cases}
1 &\text{if $A_{J,K}$ occurs},\\
0 &\text{otherwise}
\end{cases}
\end{align*}
and
\begin{align*}
Y=\sum_{J\in\mathcal{J},K\in\mathcal{K}_J}Y_{J,K},
\end{align*}
respectively.
Note that appropriately deleting $Y'\le Y$ columns from $M$ gives a $(t,n-Y',d,x)$ X-code, which implies that there exists a $(t,n-\lfloor\mathbb{E}(Y)\rfloor,d,x)$ X-code.
Hence, computing the expected value $\mathbb{E}(Y)$ gives a sufficient condition for the existence of an X-code.
Now, by linearity of expectation,
\begin{align*}
\mathbb{E}(Y)
&=\sum_{J\in\mathcal{J},K\in\mathcal{K}_J}\mathbb{E}(Y_{J,K})\\
&=\hspace{-1mm}\sum_{J\in\mathcal{J},K\in\mathcal{K}_J}\hspace{-2mm}\left(\hspace{-0.3mm}1-\hspace{-0.5mm}\sum_{l:\text{odd}}^{|J|}\binom{|J|}{l}p^l(1-p)^{|J|-l}(1-p)^x\hspace{-0.3mm}\right)^t\hspace{-0.5mm}.
\end{align*}
By Proposition \ref{Gallager}, the right-hand side of the above equality equals
\begin{align*}
\sum_{i=1}^{d}\binom{n}{x}\binom{n-x}{i}\left(1-\frac{1-(1-2p)^{i}}{2}(1-p)^x\right)^t,
\end{align*}
as desired.
\end{IEEEproof}

Theorem \ref{Co:t-lower} follows from Theorem \ref{Th:alt} as shown below.
\begin{IEEEproof}[Proof of Theorem \ref{Co:t-lower}]
Assume that $x \not= 0$.
It is routine to show that $\mathbb{E}(Y)$ in the proof of Theorem \ref{Th:alt} is bounded from above by
\begin{align*}
\mathbb{E}(Y) \leq \sum_{i=1}^{d}\binom{n}{x}\binom{n-x}{i}\left(1-p(1-p)^x\right)^t.
\end{align*}
Note that the right-hand side of the above inequality is minimized when $p = \frac{1}{x+1}$.
Since $(1-\frac{1}{x+1})^x \geq \frac{1}{e}$ for $x \geq 1$, if 
\begin{align}\label{forcomparison1}
\sum_{i=1}^{d}\binom{n}{x}\binom{n-x}{i}\left(1-\frac{1}{e(x+1)}\right)^t\le \frac{n}{2},
\end{align}
then there exists a $(t, n-\lfloor\frac{n}{2}\rfloor, d,x)$ X-code.
By rewriting $n$ as $2n$ in the above sufficient condition, there exists a $(t,n,d,x)$ X-code if
\begin{align*}
\sum_{i=1}^{d}\binom{2n}{x}\binom{2n-x}{i}\left(1-\frac{1}{e(x+1)}\right)^t\le n.
\end{align*}
By taking logarithm of both sides of the above inequality and using the fact that $\ln(1-y)\le -y$ for $0\le y<1$,
the conditions that $2n-x\geq2d$ and that
\begin{align*}
t \geq (x+1)(d+x-1)(e\ln 2)\log n + c_{d,x},
\end{align*}
where
\begin{align*}
c_{d,x} &= (x+1)(e\ln 2)\log\left(\frac{2^{d+x}}{x!(d-1)!}\right)\\
&\leq 4e(\ln16-\ln3),
\end{align*}
are sufficient for the existence of a $(t,n,d,x)$ X-code.
When $x = 0$, arguing the same way with $p =\frac{1}{2}$ gives a sufficient condition that $t \geq (d-1)\log n + d - \log((d-1)!)$.
\end{IEEEproof}

Note that, as is clear from the proof, the constant term $4e(\ln 16 - \ln 3)$ in Theorem \ref{Co:t-lower} can be strengthened considerably to elementary functions of $d$ and $x$,
giving an alternative stronger bound in closed form.

Table \ref{n1000} lists sample parameters realizable by our idea given in the proof of Theorem \ref{Co:t-lower} and the previously known simple proof of Theorem \ref{preciseY2010}.
Note that Theorems \ref{Co:t-lower} and \ref{preciseY2010} themselves are not the strongest possible bounds derivable by the ideas in their proofs
because we approximated some values to obtain cleaner, easy-to-understand bounds.
For a fair comparison, sufficient conditions similar to Inequality (\ref{forcomparison1}) in the proof of Theorem \ref{Co:t-lower} was derived by following the proof except the use of the approximation $(1-\frac{1}{x+1})^x \geq \frac{1}{e}$.
As is clearly seen, our results on the length of the shortest possible X-codes greatly improve the known ones consistently for a wide range of parameters.

\begin{table*}
\renewcommand{\arraystretch}{1.3}
\caption{\vspace{1mm}Upper bounds on the smallest $t$ for the existence of a $(t,n,d,x)$ X-code}
\label{n1000}
\centering
\begin{tabular}{cc|cc|cc|cc}
\hline
\hline
& &\multicolumn{2}{|c|}{$n=10^3$}&\multicolumn{2}{c|}{$n=10^5$}&\multicolumn{2}{c}{$n=10^7$}\\
$d$ & $x$ & Theorem \ref{Co:t-lower}\rlap{\textsuperscript{a}}   & Theorem \ref{preciseY2010}\rlap{\textsuperscript{a}} & Theorem \ref{Co:t-lower}  & Theorem \ref{preciseY2010}& Theorem \ref{Co:t-lower}  & Theorem \ref{preciseY2010}\\
\hline
1 & 1 & 29 & 49& 45 &81 & 61 &113\\
  & 2 & 95  & 150& 153 & 254 & 210 & 357\\
  & 3 & 195  & 401 & 319 & 686 & 443 & 972\\
  & 4 & 327  & 988 & 543 & 1714 & 758 & 2439\\
  & 5 & 490 & ----- & 822 & 4083 & 1154 & 5837\\
  & 6 & 681 & ----- & 1155 & 9437 & 1629 & 13547\\
\hline
3 & 1 & 76 & 90 & 124 & 154 &172 & 218\\
  & 2 & 179  & 240 & 294 & 413 & 409 & 585\\
  & 3 & 315  & 587 & 522 & 1015 & 729 & 1443\\
  & 4 & 484  & ---- & 807 & 2382 & 1131 & 3398\\
  & 5 & 683  & ----- & 1148 & 5431 & 1613 & 7771\\
  & 6 & 911  & ----- & 1543 & 12144 & 2175 & 17429\\
\hline
6 & 1 & 139  & 146 & 235 & 258 & 331 & 370\\
  & 2 & 291 & 360 & 492 & 636 & 693 & 912\\
  & 3 & 477 & 834 & 808 & 1476 & 1138 & 2118\\
  & 4 & 695 & ----- & 1180 & 3319 & 1665 & 4770\\
  & 5 & 943  & ----- & 1607 & 7320 & 2271 & 10537\\
  & 6 & -----   & ----- & 2089 & 15937 & 2958 & 22983\\
\hline
\hline
\multicolumn{8}{l}{\scriptsize\textsuperscript{a}
 The cases when the bound is weaker than the trivial one are marked by ``-----''.}\vspace{-1.1mm}\\
\end{tabular}
\end{table*}

\section{Concluding remarks}\label{sec:conclusion}
We have presented various new general bounds on the separating redundancy of a linear code.
Progress has been made both on the lower and on the upper bounds through probabilistic combinatorics and design theory.

We have also shown a striking similarity between $l$-separating parity-check matrices and X-codes, which is Proposition \ref{prop:xcodeseparation}.
This allowed us to improve known general bounds on the parameters of optimal X-codes by employing essentially the same probabilistic tool as the one we used to derive probabilistic bounds on separating redundancy.
Theorem \ref{Co:t-lower} and its slightly tighter version provided upper bounds in closed form on the shortest possible length of an X-code.
Indeed, we proved that a $(t,n,d,x)$ X-code exists for any
\[
t \geq (x+1)(d+x-1)(e\ln2)\log n + 4e(\ln16-\ln3).
\]

On the separating redundancy side, the lower bound we gave is always at least as sharp as the previously known one by Theorem \ref{lwb:volume} and quite often sharper.
It is notable that there exist linear codes that achieve our lower bound by Theorem \ref{thm:lwb} for some $l$.
Interesting examples include all MDS codes with $l = d-2$, where Theorems \ref{thm:lwb} and \ref{lwb:volume} coincide and are both achieved by the $[n,k,n-k+1]_q$ codes (see \cite{Abdel-Ghaffar:2013} for the $(d-2)$-separating redundancy of an MDS code).
Thus, any general lower bound which is at least as sharp must reduce to Theorem \ref{thm:lwb} in the achievable cases.
For this reason, it appears very difficult to give a simple and better lower bound without imposing some condition on the applicable linear codes, the range of $l$, or both.

To also bound separating redundancy from above, we refined two known approaches and proved
general bounds that are applicable even when little structural information is available other than the basic code parameters.
Through theoretical analyses and numerical examples, our theorems were shown to be much sharper than the previously known bounds in many cases.

It should be noted, however, that there is still a considerable gap between the best upper and lower bounds in general.
Although minor improvements might be possible within the same framework through, for instance, the probabilisitic method by random sampling without replacement and/or a more careful analysis on the required number of rows to fix all blemishes,
we expect that a fundamentally different approach is required to substantially improve our bounds.
In fact, the same type of mathematical difficulty appears in closely related studies that aim to find the theoretical limits of similar concepts regarding the number of rows of a special parity-check matrix such as \textit{stopping redundancy} \cite{Schwartz:2006,Han:2007,Han:2008a} for erasure channels and \textit{trapping redundancy} \cite{Laendner:2009,Tsunoda:2016a} for additive noise.

In particular, the $l$\textit{th stopping redundancy} of a linear code $\mathcal{C}$ is defined to be the smallest possible number of rows of a parity-check matrix $H$ for $\mathcal{C}$ such that
$H$ contains no stopping sets of size equal to or smaller than $l$.
Recall that $l$-separating parity-check matrices do not contain any stopping sets of size $l$ or smaller.
Thus, the separating redundancy of a linear code is at least as large as its stopping redundancy of the corresponding level $l$.
Therefore, it may not be too surprising that the idea of alterations and sample-and-modify in probabilistic combinatorics gives very strong upper bounds both on separating redundancy and on stopping redundancy.
In fact, it seems to be quite difficult to beat those probabilistic bounds unless we impose some assumption such as specifying the range of applicable parameters or focusing on a particular family of codes.
It would be an interesting combinatorial problem on its own to bound separating redundancy, stopping redundancy, and other related types of redundancy as tightly as possible in the general case.

An important question on separating redundancy we did not address is whether the error-erasure separation approach is competitive against other known decoding strategies and when the error-erasure separation approach is particularly beneficial.
While there does not seem to be a simple and complete answer to this question, it may be safe to say that it is of potential benefit for our toolbox to include an alternative approach that can handle the situations where the advantages of many other well-known approaches disappear.

To briefly discuss this aspect of error-erasure separation, consider, for instance, the fact that popular decoding strategies often take advantage of a parallelism in some way or another.
Perhaps, the simplest example is the standard approach to correcting both errors and erasures by assigning random symbols to erased positions.
In this case, if we have many independent decoders that work in parallel, we may decode the received vector in one shot by letting each independent decoder assume a different symbol pattern for the erased part.
Typical trial-based decoding methods such as Chase decoding can also significantly benefit from parallel decoding in the same manner.
However, this kind of parallelism implicitly assumes that we can freely make copies of the received vector, which is of course true in most digital communications
but may not always be the case.
Although it is too early to claim that error-erasure separation is the most competitive in some situations, 
we believe that further investigations on separating erasures from errors is of value in coding theory.

Another question we did not address is error-erasure separation for particular practical codes and situations.
In this paper, we have focused on deriving very general mathematical bounds that work for any linear code.
However, from a purely coding-theoretic point of view, we believe that it is of equal importance to investigate particular error-correcting codes in the context of error-erasure separation and whether it is practical and competitive in real communications.
Research in this direction can be found for a class of geometric low-density parity-check codes in \cite{Diao:2013}.
With the progress on separating redundancy we have made in the general case, we believe that more specialized approaches tailored to specific linear codes and research on practical implementation also deserve greater attention in future work.

Finally, we only explored the surface of the close relation between the problem of separating redundancy and that of X-codes.
For instance, it would be of interest to investigate the use of an X-code forming a parity-check matrix for a good linear code in the context of error-erasure separation.
It is expected that there remain many interesting facts to be discovered in this intersection.

\section*{Acknowledgments}
The authors thank the anonymous reviewers and Associate Editor Aditya Ramamoorthy for their careful reading of the manuscript.
Their valuable comments and constructive criticisms greatly improved the quality and readability.



\begin{IEEEbiographynophoto}{Yu Tsunoda}
(S'16) received the B.S. degree in engineering in 2016 and the M.S. degree in engineering in 2018 both from Chiba University, Chiba, Japan.
She is currently pursuing a Ph.D degree at Chiba University.
Her research interests include probabilistic combinatorics, coding theory, and their interactions.
\end{IEEEbiographynophoto}

\begin{IEEEbiographynophoto}{Yuichiro Fujiwara}
(M'10) received the B.S. and M.S. degrees in mathematics from Keio University, Japan,
and the Ph.D. degree in information science from Nagoya University, Japan.

He was a JSPS postdoctoral research fellow with Tsukuba University, Japan, Michigan Technological University,
and California Institute of Technology. He is currently an Associate Professor with the Division of Mathematics and Informatics, Chiba University, Japan.

Dr.\ Fujiwara's research interests include combinatorics and its interaction with computer science, quantum information science, and electrical engineering,
with particular emphasis on combinatorial design theory, probabilistic combinatorics, algebraic coding theory, and quantum information theory.
\end{IEEEbiographynophoto}

\begin{IEEEbiographynophoto}{Hana Ando}
received her M.S. degree in Engineering from Chiba University, Chiba, Japan, in 2018.
She is currently a system engineer at Nomura Research Institute, Ltd in Japan.
\end{IEEEbiographynophoto}

\begin{IEEEbiographynophoto}{Peter Vandendriessche}
obtained the B.S. in Mathematics in 2008, and
simultaneously the M.S. in Mathematics and M.S. in Computer Science in
2010, both from Ghent University. In 2014, he completed his PhD in
Mathematics, also at Ghent University, where he is currently a postdoctoral
researcher on a grant by the FWO-Vlaanderen.
\end{IEEEbiographynophoto}

\end{document}